



\magnification\magstep1
\vsize=24.0truecm
\baselineskip14pt
\def\hatt{\widehat}
\def\dell{\partial}
\def\tilda{\widetilde}
\def\eps{\varepsilon}
\def\onebyn{\hbox{$1\over n$}}
\def\half{\hbox{$1\over2$}}
\font\csc=cmcsc10
\def\cstok#1{\leavevmode\thinspace\hbox{\vrule\vtop{\vbox{\hrule\kern1pt
        \hbox{\vphantom{\tt/}\thinspace{\tt#1}\thinspace}}
        \kern1pt\hrule}\vrule}\thinspace} 
\def\firkant{\cstok{\phantom{$\cdot$}}} 

\overfullrule=0pt

\def\dd{{\rm d}}
\def\tr{{\rm t}}

\centerline{\bf On inference in parametric survival data models} 

\bigskip 
\centerline{\bf Nils Lid Hjort} 

\medskip
\centerline{\sl University of Oslo and Norwegian Computing Centre}
\smallskip
\centerline{\sl October 1991}

\medskip 
{{ \smallskip\narrower\baselineskip12pt 
\noindent {\sl Department of Mathematics and Statistics, 
University of Oslo, N--0316 Oslo, Norway; 
and Norwegian Computing Centre, N--0314 Oslo 3, Norway}

\medskip\noindent 
{\csc Summary}. 
The usual parametric models for survival data are of the following form.
Some parametrically specified hazard rate $\alpha(s,\theta)$ is 
assumed for possibly censored random life times $X_1^0,\ldots,X_n^0$;
one observes only $X_i=\min\{X_i^0,c_i\}$ and $\delta_i=I\{X_i^0\le c_i\}$
for certain censoring times $c_i$ that either are given or come from some
censoring distribution. 
We study the following problems:
What do the maximum likelihood estimator and other estimators 
really estimate when the true hazard rate $\alpha(s)$ is different from the 
parametric hazard rates? 
What is the limit distribution of an estimator
under such outside-the-model circumstances? 
How can traditional model-based analyses be made model-robust? 
Does the model-agnostic viewpoint invite 
alternative estimation approaches?
What are the consequences
of carrying out model-based and model-robust bootstrapping?
How do theoretical and empirical influence functions
generalise to situations with censored data?  
How do methods and results carry over to more complex models for 
life history data like regression models and Markov chains? 

\smallskip\noindent 
{\csc Key words:} \sl
agnostic parameter estimation;
censored data; 
distance measures;
hazard regression;
incorrect model; 
influence function;
maximum likelihood; 
parametric and nonparametric bootstrapping \smallskip}} 

\bigskip
\centerline{\bf 1. Introduction}

\medskip 
This paper is about aspects of maximum likelihood 
and related estimation methods 
applied to parametric survival data models.
The aspects we shall care about include 
large-sample behaviour when the parametric model is a nonperfect
approximation to the true model;
distance measures from true to parametric model; 
model-based and model-robust estimation of the approximate
covariance matrix of the maximum likelihood estimator; 
measures of influence; 
natural alternative estimation procedures suggested by the 
agnostic point of view; 
model-based and model-robust ways of bootstrapping;
and similar questions for hazard rate regression models.
Indeed, 
Section 2 studies limit behaviour of the maximum 
likelihood estimator when the parametric model is incorrect,
Section 3 finds influence functions under censoring, 
and in Section 4 the general methods are used 
to assess the behaviour of various bootstrapping schemes. 
The apparatus developed in Sections 3 and 4 
can be used to prove some known results anew,
and should be useful also in other survival data models 
and for other estimators than the maximum likelihood one. 
Some new estimation methods are discussed in Section 5,
and Section 6 treats two regression models for hazard rates.
Complementary 
remarks are offered in the final Section 7.

A recurrent theme underlying our article is the point of view that 
(i) parametric models are usually incorrect, 
(ii) that estimation and inference in parametric models 
nevertheless can be a useful enterprise,
(iii) provided the statistician knows what she is doing.
Even statisticians admit (i). 
Traditional and valid arguments favouring (ii) include matters of
sample size versus nonparametrics and the value of 
simplifying and synthesising to aid understanding
of complex phenomena.   
The following reasoning also supports (ii) and pertains to the present
paper. We view a parametric estimation procedure 
as an attempt to find the best fitting
or most appropriate parametric approximant to the more elusive
true model. An estimator for the parameter vector $\theta$ 
will typically be consistent for a certain $\theta_0$ that is
most appropriate, or least false, in the sense of minimising 
a suitable distance measure between true model and parametric model. 
Accordingly estimating the 
least false parameter is a meaningful statistical operation,
even outside model conditions (i.e.~even if the minimum distance
is positive), provided only that the distance measure 
itself is reasonable. 
Regarding (iii) above, 
as far as the first order large sample consequences of an 
incorrect parametric model is concerned, 
the single technical complication 
will be seen to be a different expression for the 
limiting covariance matrix of the estimators. 
A consistent estimator for this more general covariance matrix 
can be constructed explicitly, or approximated by appropriate resampling, 
or reached as a by-product of empirical influence functions. 

Different estimation methods may correspond to different distance
measures and thus different least false parameters. 
It often enhances one's understanding of an estimation procedure
to view it in this light, i.e.~by exhibiting the
accompanying distance measure between truth and approximating model. 
A case in point is the maximum likelihood method which amounts 
to minimising an estimate of the Kullback--Leibler distance from
true density to approximating density. 
Of course this agnostic point of view can be the explicit motivation
for some estimators in the first place; 
an empirical counterpart can be constructed for a given distance measure
and then be minimised for the given data.
In Section 2 the distance measure underlying maximum likelihood
estimation in survival data models is exhibited, 
thereby generalising the Kullback--Leibler distance, 
and a couple of alternative distance measures and estimators 
are discussed in Section 5.  
Distance measures behind standard methods
for estimation in Cox regression models are found in Section 6.
These involve both the censoring distribution and the 
covariate distribution. 

The results of this paper give precise statistical substance 
to fitting and analysing data with a wrong model, and suggest that 
it can even be fruitful. This is not to say that one shouldn't
assess the adequacy of one's model or compare different natural
models; one should indeed, and general methods for doing this 
can be found in Hjort (1990). But the agnostic point of view
and results under such is meant to free statisticians from the irongrip
of that part of traditional methodology which has 
`the parametric model is assumed to be absolutely correct' 
as basic assumption. This should have some pragmatic value 
as well, since practitioners often try out a variety of models 
while knowing that neither of them is likely to be quite correct. 
The theory developed below gives a recipe for bettering this 
practice by using corrected approximate covariance matrices for
the estimators. 

The points of view expressed above are not entirely new, 
but relatively few publications have discussed behaviour of
model-derived estimates under fixed alternative conditions.
Variations of the basic result (1.3) below has 
for example appeared a couple of times in various contexts, 
and sometimes rather implicitly, 
see Cox (1962), Huber (1967), Chibisov (1973), 
and Reeds (1978) for early examples and 
White (1982), Hjort (1986a, 1986b, 1988), 
and Linhart and Zucchini (1986)
for recent ones in different settings. 
For related views for stochastic process models,
see McKeague (1984) and Barndorff-Nielsen and S\o rensen (1991).
Finally Lin and Wei (1989) also worked with misspecified 
Cox regression models, and obtained results similar to those
of Section 6B below, independently of the present narrator. 
Our results are more general in that we exhibit 
the distance function that Cox estimation corresponds to,
find the theoretical and empirical influence functions, 
and also explore parametric Cox regression. 

One can also usefully define and study situations where the amount of 
misspecification is moderate. This is done on a general basis in
Hjort (1991a). Included there is a result which says that 
it is actually advantageous, in terms of precision of estimators,
to stick to a given model even when it is moderately incorrect,
and the precise `tolerance radius' around the model against
various types of model departures is also found. 

The remainder of this section is a concise treatment
of the simpler non-censored i.i.d.-case. 
It is included here since the viewpoint and results 
do not appear to be well known, and  
since our results perhaps will be easiest to understand and appreciate
when compared to corresponding statements 
for this simpler classical framework. 

Let $X_1,\ldots,X_n$ be independent
from some unknown distribution $F$ with density $f$, 
and suppose the data are to be fitted
to some $p$-dimensional parametric family of densities 
$\{f_\theta\colon\theta\in\Theta\}$.
Where notationally convenient we shall write $f(x,\theta)$ instead of 
$f_\theta(x)$ and so on. 
Note that we do not assume the true $f$ 
to belong to the parametric class,
unlike what is typically the case in 
textbook treatments of this problem.
The maximum likelihood estimator $\hatt\theta$ maximises
the observed likelihood $L_n(\theta)$ 
w.r.t.~the parameter. Since the simple average 
$n^{-1}\log L_n(\theta)$ tends to 
$E_F\log f_\theta(X)=\int f\log f_\theta\,\dd x$ in probability
$\hatt\theta$ intuitively aims at becoming close to the parameter
value $\theta_0$ that maximises this expression, or,
equivalently, minimises the Kullback--Leibler distance
$$d[f,f_\theta]=\int f(x)\log\{f(x)/f_\theta(x)\}\,\dd x \eqno(1.1)$$
from true model to parametric model.
(Although it is not symmetric in its arguments 
we shall term it and its later generalisations `distances'.)  
We think of $\theta_0=\theta_0(F)$, 
which is indeed uniquely defined in most
cases, as the {\it least false} or {\it most fitting}
parameter value. 

We summarise below the behaviour of $\hatt\theta$ for large $n$ under the
present outside-the-model circumstances. The arguments needed to prove
the results can be seen as more careful versions of the `traditional ones' 
that are used under model circumstances (see e.g.~Lehmann, 1983, Ch.~6). 
Consider the $p$-vector $U_n$ of first order derivatives and the 
$p\times p$-matrix $I_n$ of second order derivatives of 
$n^{-1}\log L_n(\theta)$.
$\hatt\theta$ is a solution to the maximum likelihood equations 
$U_n(\theta)=0$, so by Taylor expansion 
$0=U_n(\hatt\theta)=U_n(\theta_0)+I_n(\tilda\theta)(\hatt\theta-\theta_0)$,
which leads to
$$\sqrt{n}(\hatt\theta-\theta_0)=\{-I_n(\tilda\theta)\}^{-1}
	\,\sqrt{n}\,U_n(\theta_0), \eqno(1.2)$$
in which $\tilda\theta$ lies somewhere between $\theta_0$ and $\hatt\theta$.
Two matrices therefore determine the limit distribution:
the limit $J=J(F,\theta_0)$ of $-I_n(\theta_0)$, 
obtained by the law of large numbers,
and the covariance matrix $K=K(F,\theta_0)$ of $\sqrt{n}\,U_n(\theta_0)$, 
obtained from the central limit theorem. More precisely, 
$$J=-\int {\dell^2\log f(x,\theta_0) 
	\over \dell\theta\dell\theta}\,\dd F(x) 
	\quad {\rm and} \quad 
K=\int\Bigl({\dell\log f(x,\theta_0)
	\over \dell\theta}\Bigr)
	\Bigl({\dell\log f(x,\theta_0)\over 
		\dell\theta}\Bigr)^\tr\,\dd F(x).$$
Natural estimators for these $p\times p$ matrices are
$\hatt J=J(\hatt F,\hatt\theta)$ and 
$\hatt K=K(\hatt F,\hatt\theta)$, 
that is
$$\hatt J=-{1\over n}\sum_{i=1}^n{\dell^2\log f(X_i\hatt\theta)
	\over \dell\theta\dell\theta},
	\,
  \hatt K=K(\hatt F,\hatt\theta)={1\over n}\sum_{i=1}^n
	\Bigl({\dell\log f(X_i,\hatt\theta)\over \dell\theta}\Bigr)
	\Bigl({\dell\log f(X_i,\hatt\theta)\over \dell\theta}\Bigr)^\tr.$$
Here $\hatt F$ is the empirical distribution which places weight 
$1/n$ on each data point.
 	
\smallskip
{\csc Result.} {{\sl 
Under traditional regularity conditions $\hatt\theta$ is consistent
for the least false parameter $\theta_0$. Furthermore,
$$\sqrt{n}(\hatt\theta-\theta_0)\rightarrow_d 
	J^{-1}N_p\{0,K\}
	=N_p\{0,J(F,\theta_0)^{-1}K(F,\theta_0)
					J(F,\theta_0)^{-1}\}, \eqno(1.3)$$
and $\hatt J$ and $\hatt K$ are consistent estimators for $J$ and $K$.}}

\smallskip
The result (1.3) is the appropriate generalisation of the classical
textbook result, in which $f(x)=f(x,\theta_0)$ is assumed, and where
it is easy to show that the two matrices are equal,
$$J(F_\theta,\theta)=K(F_\theta,\theta). \eqno(1.4)$$ 
%
%
We can now distinguish between model-based and model-robust 
inference about $\theta_0$. In the first case 
$\theta_0$ is true, and one 
uses ${\tilda J}^{-1}/n$ as the estimate of the
covariance matrix for $\hatt\theta$,
where $\tilda J$ could be either $J(\hatt F,\hatt\theta)$
or $J(F(.,\hatt\theta),\hatt\theta)$.
In the second case $\theta_0$ has the wider interpretation of 
being merely most fitting, and one uses 
${\hatt J}^{-1}\hatt K{\hatt J}^{-1}/n$ instead.

\smallskip
{\csc Example 1.1.} Suppose nonnegative data are fitted to the exponential
distribution with density $f_\theta(x)=\theta\exp(-\theta x)$. Then 
$d[f,f_\theta]=\int_0^\infty f(x)\log f(x)\,\dd x
-\int_0^\infty(\log\theta-\theta x)f(x)\,\dd x$ is minimised for
the least false parameter $\theta_0=1/\mu(F)$, where $\mu(F)=E_F X$.
One finds $J=1/\theta_0^2$ and $K={\rm Var}_F X=\sigma^2(F)$.
The model-based asymptotic variance of $\hatt\theta=1/\hatt\mu$
is $n^{-1}\theta_0(F)^2$, estimated by $n^{-1}\hatt\theta^2$, 
whereas the model-robust version is $n^{-1}\sigma^2(F)\theta_0(F)^4$,
estimated by $n^{-1}\hatt\sigma^2\hatt\theta^4$. \firkant

\smallskip
Next turn attention to bootstrapping. Model-based bootstrapping consists
of drawing samples $X_1^*,\ldots,X_n^*$ from the parametrically
estimated $F(.,\hatt\theta)$, 
and computing bootstrap estimates 
$\hatt\theta^*=\hatt\theta(X_1^*,\ldots,X_n^*)$. Nonparametric or
model-robust bootstrapping on the other hand samples $X_i^*$'s
from $\hatt F$. The (first-order) large sample behaviour of $\hatt\theta^*$ 
can be analysed and characterised by the methods already used. 
Think of $\theta_0={\rm ml}(F)$, 
the maximiser of $\int\log f_\theta(x)\,\dd F(x)$, 
assuming this to be unique, 
as a functional operating on the space of distributions. 
Observe that both ${\rm ml}(\hatt F)$ and ${\rm ml}(F(.,\hatt\theta))$
are equal to $\hatt\theta$. 
By (1.2) and (1.3) we have 
$$\sqrt{n}\{{\rm ml}(\hatt F)-{\rm ml}(F)\}\doteq_d
	J(F,{\rm ml}(F))^{-1}{1\over \sqrt{n}}\sum_{i=1}^n
	{\dell\log f(X_i,{\rm ml}(F))\over \dell\theta}, \eqno(1.5)$$
where $U_n\doteq_d V_n$ means that $U_n-V_n$ tends to zero in probability.
More precise information can be gathered using methods 
presented in Section 4.

Consider first parametric bootstrapping, 
which uses $\hatt\theta^*$ computed from $F(.,\hatt\theta)^*$, say,
the empirical distribution of $X_i^*$'s from $F(.,\hatt\theta)$. 
Then, conditionally on the observed data,   
$$\eqalign{
\sqrt{n}(\hatt\theta_{\rm pb}^*-\hatt\theta)
	&=\sqrt{n}\{{\rm ml}(F(.,\hatt\theta)^*)-{\rm ml}(F(.,\hatt\theta))\}
	 \doteq_d J(F(.,\hatt\theta),\hatt\theta)^{-1}
	 {1\over \sqrt{n}}\sum_{i=1}^n
		{\dell\log f(X_i^*,\hatt\theta) \over \dell\theta} \cr
	&\doteq_d J(F(.,\hatt\theta),\hatt\theta)^{-1}
		N_p\{0,K(F(.,\hatt\theta),\hatt\theta)\} 
	 =N_p\{0,J(F(.,\hatt\theta),\hatt\theta)^{-1}\}. \cr} \eqno(1.6)$$
Correspondingly, for nonparametric bootstrapping one has 
$$\eqalign{
\sqrt{n}(\hatt\theta_{\rm nb}^*-\hatt\theta)
	 &=\sqrt{n}\{{\rm ml}(\hatt F^*)-{\rm ml}(\hatt F)\}
	  \doteq_d J(\hatt F,\hatt\theta)^{-1}
	  {1\over \sqrt{n}}\sum_{i=1}^n
		{\dell\log f(X_i^*,\hatt\theta) \over \dell\theta} \cr
	&\doteq_d J(\hatt F,\hatt\theta)^{-1}
		N_p\{0,K(\hatt F,\hatt\theta)\} 
	 =N_p\{0,\hatt J^{-1}\hatt K\hatt J^{-1}\}. \cr} \eqno(1.7)$$
Several conclusions can be drawn from this. 
First, the nonparametric bootstrap always works, in the large sample 
first order sense, in that the bootstrap distribution always mimics 
the true distribution, even when the parametric model is incorrect;
the distribution of $\sqrt{n}(\hatt\theta_{\rm nb}^*-\hatt\theta)$
tends with probability one to the same as does 
$\sqrt{n}(\hatt\theta-\theta_0)$, cf.~(1.3). 
Secondly, the parametric bootstrap only works when the model is correct,
otherwise it does not reflect the real sampling variability. 
Thirdly, we should note that the sampling variability of
$\hatt\theta^*_{\rm nb}$ is typically much larger than that of 
$\hatt\theta^*_{\rm pb}$. 
This is related to the observation that 
{\it if} the model happens to be correct,
then both $\hatt J^{-1}\hatt K\hatt J^{-1}$ and $\hatt J^{-1}$ estimate
the same quantity, namely the asymptotic covariance matrix
of $\sqrt{n}(\hatt\theta-\theta_0)$, 
but the first is less stable than the second. 

In situations where interest centres on another parameter 
$\mu=\mu(\theta)$ the discussion here applies to 
$\hatt\mu=\mu(\hatt\theta)$ and $\hatt\mu^*=\mu(\hatt\theta^*)$ instead.

\smallskip
{\csc Example 1.2.} Let $\hatt V_{\rm nb}$ and $\hatt V_{\rm pb}$ be
the bootstrap estimates of the variance of $\hatt\theta$ in the exponential
situation treated above. Then it can be shown that
$${{\rm Var}\{\hatt V_{\rm pb}\}\over {\rm Var}\{\hatt V_{\rm nb}\}} 
	\doteq  
  {{\rm Var}\{\hatt\theta^2/n\} \over 
	{\rm Var}\{\hatt\theta^4\hatt\sigma^2/n\}}
	\doteq 
	{4\theta_0^4/n \over 8\theta_0^4/n}={1\over 2} $$
if the exponential model prevails. 
See also further comments, examples, 
and amendments in Hjort (1988). \firkant

\bigskip
\centerline{\bf 2. Theory for incorrectly specified 
		parametric survival data models}

\medskip 
Suppose $X_1^0,\ldots,X_n^0$ are lifetimes for $n$ individuals drawn
from a homogeneous population with underlying hazard rate
$\alpha(s)=f(s)/F[s,\infty)$ for $s\ge0$. 
Suppose that one observes only 
$X_i=\min\{X_i^0,c_i\}$ and $\delta_i=I\{X_i^0\le c_i\}$,
where the censoring variables $c_i$ 
are independent of the lifetimes and   
come from some censoring distribution $G$. 
A parametric model is proposed of the type
$\alpha(s)\approx\alpha_\theta(s)=\alpha(s,\theta)$. 
In this section the large-sample
properties of the maximum likelihood estimator outside model conditions
are derived, parallelling the treatment of the traditional 
non-censored type problem in Section 1.

The treatment below extends that of Borgan (1984) and Hjort (1986a).
The mathematical techniques needed to derive results involve
central limit theorems and inequalities for martingales and
integrals of previsible functions with respect to martingales.
The necessary technicalities resemble those thoroughly presented in 
Andersen and Gill (1982), Borgan (1984), 
Andersen and Borgan (1985), Hjort (1986a),
and in the recent book Andersen, Borgan, Gill and Keiding (1992). 
This allows us to skip most of the formal details here. 
New proofs of some of the older results 
can also be constructed as a 
by-product of the general machinery of influence functions and 
differentiable functionals developed in Sections 3 and 4 below. 

We must start by defining the maximum likelihood estimator. Introduce 
the counting process $N$, the at-risk process $Y$, and the associated
martingale $M$ by
$$N(t)=\sum_{i=1}^nI\{X_i\le t,\delta_i=1\}, \quad
	Y(t)=\sum_{i=1}^nI\{X_i\ge t\}, \quad 
	M(t)=N(t)-\int_0^tY(s)\alpha(s)\,\dd s. \eqno(2.1)$$
Notice that $M$ employs the true hazard rate $\alpha(s)$ 
rather than some $\alpha(s,\theta_0)$.
We could have subscripted $N$, $Y$, $M$ with $n$ to indicate 
their dependence upon sample size $n$,
but this would later lead to overburdening since we 
also shall need the individual processes $N_i$, $Y_i$, $M_i$. 
With conditions about the censoring mechanism much weaker than
the random censorship assumption used here the likelihood 
can be written 
$$L_n(\theta)=\exp\Bigl[\int_0^T\bigl\{\log\alpha(s,\theta)\,\dd N(s)
	-Y(s)\alpha(s,\theta)\,\dd s\bigr\}\Bigr],$$
where $[0,T]$ is the time interval over which the processes are observed.
We will assume $T$ finite to get certain martingale arguments below
easily through, but extension to the full half-line is possible with
appropriate extra conditions. 
Among the important properties of $M$ is the fact that
$W_n=\int_0^tH_n(s)\,\dd M(s)/\sqrt{n}$ converges in distribution to
$W=\int_0^th(s)\,dV(s)$, provided $H_n$ is previsible (the value of
$H_n(s)$ is known already at time $s-$) and converges uniformly, 
in probability, to a deterministic function $h$.
Here $V$ is a Gau\ss ian zero-mean process
with independent increments and ${\rm Var}\,dV(s)=y(s)\alpha(s)\,\dd s$,
and $y(s)$ is the limit in probability of $Y(s)/n$,  
namely
$$y(s)={\rm Pr}\{X_i\ge s\}={\rm Pr}\{X_i^0\ge s,c_i\ge s\}
	=F[s,\infty)G[s,\infty). \eqno(2.2)$$
See Andersen and Borgan (1985), for example. 
Note that $W$ is normal with mean zero 
and variance $\int_0^th^2y\alpha\,\dd s$.

Consider first 
$$\eqalign{
{1\over n}\log L_n(\theta)&={1\over n}\int_0^T
	\bigl\{\log\alpha(s,\theta)\,\dd N(s)-Y(s)\alpha(s,\theta)\,\dd s\bigr\} \cr
	&={1\over n}\int_0^T\bigl\{\log\alpha_\theta (\dd M+Y\alpha\,\dd s)
		-Y\alpha_\theta \dd s\bigr\} 
	 \rightarrow_p \int_0^T 
	  y(\alpha\log\alpha_\theta-\alpha_\theta)\,\dd s. \cr}$$
Maximising $L_n(\theta)$ should therefore in the end amount to maximising
the right hand expression here, 
which is the same as minimising the distance 
$$d[\alpha,\alpha_\theta]
	=\int_0^T y\bigl\{\alpha(\log\alpha-\log\alpha_\theta)
	 -(\alpha-\alpha_\theta)\bigr\}\,\dd s \eqno(2.3)$$
from the true model to the approximating parametric model. 
The theorem below makes this argument rigorous. 
Observe that $d[\alpha,\alpha_{\theta_0}]$ 
is always nonnegative and is zero only if
$\alpha(s)=\alpha(s,\theta_0)$ a.e.~on $[0,T]$, 
in which case $\theta_0$ indeed is the ``true'' parameter. 
In general we can only reckon with a {\it least false} 
parameter value $\theta_0$ which minimises (2.3). 
Observe that the value of $\theta_0$ may depend upon the censoring
distribution through $y(s)=F[s,\infty)G[s,\infty)$. Note also that  
(2.3) properly generalises the Kullback--Leibler distance (1.1). 
See Remark 7A and 
Section 5B. 

\smallskip
{\csc Theorem 2.1.} {{\sl 
Suppose that there is a unique minimiser $\theta_0$ of (2.3),
and that it is an innerpoint in the parameter space;
that $\alpha(s,\theta)$ is three times 
differentiable in a neighbourhood $N(\theta_0)$ of $\theta_0$; 
that these functions are bounded over $[0,T]\times N(\theta_0)$;
that $\alpha(s)$ and $\alpha(s,\theta_0)$ are bounded away from zero as 
$s$ runs from 0 to $T$; and finally that
the $J$ matrix appearing below is positive definite.
[Somewhat weaker sufficient conditions can be put up in the style of 
Borgan (1984, Section 4; note the corrigendum p.~275).]  
Then the maximum likelihood estimator $\hatt\theta$ is
consistent for the least false parameter $\theta_0$.
Consider matrices
$J=J(\alpha,y,\theta_0)$ and $K=K(\alpha,y,\theta_0)$ 
defined as follows:
$$\eqalign{
J&=\int_0^Ty(s)\Bigl[\psi(s,\theta_0)\psi(s,\theta_0)^\tr
	 \alpha(s,\theta_0)-D\psi(s,\theta_0)\{\alpha(s)
		-\alpha(s,\theta_0)\}\Bigr]\,\dd s, \cr
K&=\int_0^T\Bigl[y(s)\psi(s,\theta_0)\psi(s,\theta_0)^\tr\alpha(s)
	 +\bigl\{\psi(s,\theta_0)E(s)^\tr
		+E(s)\psi(s,\theta_0)^\tr\bigr\}			
		\alpha(s,\theta_0)\Bigr]\,\dd s, \cr}$$
in which 
$\psi(s,\theta)=\dell\log\alpha(s,\theta)/\dell\theta$, 
$D\psi(s,\theta)=\dell^2\log\alpha(s,\theta)/\dell\theta\dell\theta$, 
and 
$E(s)=\int_0^sy(t)\allowbreak\psi(t,\theta_0)\allowbreak\{\alpha(t)
			-\alpha(t,\theta_0)\}\,\dd t$
(in particular, $E(0)=E(T)=0$, by the nature of $\theta_0$).
Then 
$$\sqrt{n}(\hatt\theta-\theta_0)\rightarrow_dJ^{-1}N_p\{0,K\}
	=N_p\{0,J(\alpha,y,\theta_0)^{-1}K(\alpha,y,\theta_0)
			J(\alpha,y,\theta_0)^{-1}\}.$$ }}
\quad 
{\csc Proof:} 
It was indeed shown already in Hjort (1986a) that $\hatt\theta$ is
consistent for this most fitting parameter $\theta_0$.
There is even almost sure convergence in the present 
random censorship situation, a fact used in Section 4. 

Next turn to the limit distribution of $\hatt\theta$. 
The idea is to consider the vector $U_n$ of first order partial derivatives 
and the matrix $I_n$ of second order partial derivatives of 
$n^{-1}\log L_n$ and apply (1.2) again, in this more difficult situation. 
One finds 
$$\eqalign{
U_n(\theta)&={1\over n}\int_0^T\psi(s,\theta)\bigl\{\dd N(s)
	-Y(s)\alpha(s,\theta)\,\dd s\bigr\} \cr
	&={1\over n}\int_0^T\psi_\theta\bigl\{\dd M
			+Y(\alpha-\alpha_\theta)\,\dd s\bigr\} 
	 \rightarrow_p\int_0^T y\psi_\theta (\alpha-\alpha_\theta)\,\dd s \cr}$$
and 
$$\eqalign{
I_n(\theta)&={1\over n}\int_0^T\bigl[D\psi(s,\theta)
	\{\dd N(s)-Y(s)\alpha(s,\theta)\,\dd s\}
	-\psi(s,\theta)Y(s)\alpha(s,\theta)\psi(s,\theta)^\tr\,\dd s\bigr] \cr
	&\rightarrow_p 
	\int_0^T y\bigl\{D\psi(.,\theta)(\alpha-\alpha_\theta)\,\dd s
		 -\psi_\theta\psi_\theta^\tr\alpha_\theta\,\dd s\bigr\}. \cr}$$
In particular $-I_n(\theta_0)$ tends to the $J$ matrix in probability.
Note next that $U_n(\theta)$ tends to 
zero when $\theta=\theta_0$. Furthermore,
$$\sqrt{n}\,U_n(\theta_0)=\int_0^T\psi(s,\theta_0)\bigl[\dd M(s)/\sqrt{n} 
		+\sqrt{n}\{Y(s)/n-y(s)\}\{\alpha(s)
			-\alpha(s,\theta_0)\}\,\dd s\bigr]. $$
Here $V_n=M/\sqrt{n}$ has a limit process $V$ described before (2.2) 
and $Z_n=\sqrt{n}(Y/n-y)$ converges in distribution to a Gau\ss ian
zero-mean process $Z(.)$, in the function space $D[0,T]$ of left-continuous
functions with right hand limits, by the theory presented for example in 
Billingsley (1968, Section 13). 
One has 
$${\rm cov}\{Z_n(s),Z_n(t)\}
	={\rm cov}\bigl[I\{X_i^0\ge s,c_i\ge s\},
			I\{X_i^0\ge t,c_i\ge t\}\bigr]
		=y(s\vee t)-y(s)y(t),$$
writing $s\vee t=\max\{s,t\}$. Also, if $N_i$, $Y_i$, $M_i$ are the
counting process, at risk process, and martingale for individual no.~$i$,
then 
$${\rm cov}\{dV_n(s),Z_n(t)\}
	={\rm cov}\{\dd M_i(s),Y_i(t)\}
	=E\{\dd N_i(s)-Y_i(s)\alpha(s)\,\dd s\}Y_i(t)$$
can be seen to equal $-\alpha(s)\,\dd s\,y(t)$ for $s<t$ and 0 for $s\ge t$.
These are also expressions for ${\rm cov}\{Z(s),Z(t)\}$ 
and ${\rm cov}\{dV(s),Z(t)\}$. That indeed
$(V_n,Z_n)\rightarrow_d(V,Z)$ in $D[0,T]\times D[0,T]$ and 
$$\sqrt{n}\,U_n(\theta_0)\rightarrow_d\int_0^T\psi(s,\theta_0)\bigl[dV(s)
		+Z(s)\{\alpha(s)-\alpha(s,\theta_0)\}\,\dd s\bigr]
	={\rm one}+{\rm two} \eqno(2.4)$$
hold, where an expression for $K={\rm VAR}\{{\rm one}+{\rm two}\}$
for this necessarily Gau\ss ian limit vector 
is derived below, can be shown combining function space asymptotics
from Billingsley (1968) and Andersen and Borgan (1985). 

To find $K$, observe first that
$${\rm VAR}\{{\rm one}\}=\int_0^Ty(s)\psi(s,\theta_0)\psi(s,\theta_0)^\tr
	\alpha(s)\,\dd s.$$
Write $\Delta(s)=\alpha(s)-\alpha(s,\theta_0)$ for the difference between
true hazard and most fitting hazard. Then
$$\eqalign{
{\rm VAR}\{{\rm two}\}&=\int_0^T\int_0^T\psi(s,\theta_0)\psi(t,\theta_0)^\tr
	\Delta(s)\Delta(t)\{y(s\vee t)-y(s)y(t)\}\,\dd s\,\dd t \cr
	&=\int_0^T\int_0^t\bigl\{\psi(s,\theta_0)\psi(t,\theta_0)^\tr
		+\psi(t,\theta_0)\psi(s,\theta_0)^\tr\bigr\}
		\Delta(s)\Delta(t)y(t)\,\dd s\,\dd t, \cr}$$
since $\int_0^Ty(s)\psi(s,\theta_0)\Delta(s)\,\dd s$ is zero. 
Finally we need
$$\eqalign{
E\bigl[\{{\rm one}\}\{{\rm two}\}^\tr&+\{{\rm two}\}\{{\rm one}\}^\tr\bigr] \cr
	&=-\int_0^T\int_0^t\bigl\{\psi(s,\theta_0)\psi(t,\theta_0)^\tr
		+\psi(t,\theta_0)\psi(s,\theta_0)^\tr\bigr\}
		\alpha(s)\Delta(t)y(t)\,\dd s\,\dd t. \cr}$$
Write $\alpha(s)=\alpha(s,\theta_0)+\Delta(s)$ here, 
and find that some terms luckily cancel each other out: 
$$\eqalign{
K&=\int_0^Ty(s)\psi(s,\theta_0)\psi(s,\theta_0)^\tr\alpha(s)\,\dd s \cr
 & \qquad -\int_0^T\int_0^t\bigl[\psi(s,\theta_0)\psi(t,\theta_0)^\tr
		+\psi(t,\theta_0)\psi(s,\theta_0)^\tr\bigr]
		 \alpha(s,\theta_0)\Delta(t)y(t)\,\dd s\,\dd t. \cr}$$
The alternative formula 
given in the theorem follows upon clever integration by parts. \firkant


\smallskip
Suppose for a minute that the model is in fact true, so that
$\alpha(s)=\alpha(s,\theta_0)$. Then $J$ and $K$ agree, and there
is an identity
$$J(\alpha_\theta,y,\theta)=K(\alpha_\theta,y,\theta)
	=\int_0^Ty(s)\psi(s,\theta)\psi(s,\theta)^\tr\alpha(s,\theta)\,\dd s
						\eqno(2.5)$$
which generalises (1.4). The model-based statement 
$\sqrt{n}(\hatt\theta-\theta_0)\rightarrow_d N_p\{0,J^{-1}\}$ 
was one of the main results of Borgan (1984), and further discussion,
including matters of optimality, can be found in Hjort (1986a).

To carry out valid large-sample inference about the most fitting 
parameter $\theta_0$, for example
setting an approximate confidence interval for one of the parameter
components, one needs a consistent estimator for the asymptotic covariance
matrix. Estimators for $J$ and $K$ can be constructed in several ways.
The most natural estimators come forward when we express them as 
functions of the true cumulative hazard $A(.)=\int_0^.\alpha(s)\,\dd s$, 
the limiting at risk proportion $y(.)$, and the parameter $\theta_0$,
and then insert consistent estimators $\hatt A(.)=\int_0^. \dd N(s)/Y(s)$,
$\hatt y(.)=Y(.)/n$, and $\hatt\theta$ for these. This leads to
$$\hatt J=\int_0^T{Y(s)\over n}\psi(s,\hatt\theta)\psi(s,\hatt\theta)^\tr
	\alpha(s,\hatt\theta)\,\dd s
	-\int_0^T{Y(s)\over n}D\psi(s,\hatt\theta)
	\Bigl\{{\dd N(s)\over Y(s)}
        -\alpha(s,\hatt\theta)\,\dd s\Bigr\},\eqno(2.6)$$
and three different expressions for $\hatt K$:
$$\eqalign{
\hatt K&=\int_0^T{Y(s)\over n}
		\psi(s,\hatt\theta)\psi(s,\hatt\theta)^\tr{\dd N(s)\over Y(s)}
	+\int_0^T\bigl\{\psi(t,\hatt\theta)\hatt E(t)^\tr
		+\hatt E(t)\psi(t,\hatt\theta)^\tr\bigr\}
			\alpha(t,\hatt\theta)\,\dd t, \cr
	&=\int_0^T\psi(s,\hatt\theta)\psi(s,\hatt\theta)^\tr{\dd N(s)\over n} \cr
	 &\quad -\int_0^T\int_0^t\bigl\{\psi(s,\hatt\theta)\psi(t,\hatt\theta)^\tr
		+\psi(t,\hatt\theta)\psi(s,\hatt\theta)^\tr\bigr\}
	  \alpha(s,\hatt\theta)\,\dd s\,\Bigl\{{\dd N(t)\over n}
		-{Y(t)\over n}\alpha(t,\hatt\theta)\,\dd t\Bigr\} \cr
	&={1\over n}\sum_{i=1}^n
	  \bigl\{\psi(x_i,\hatt\theta)\delta_i
			-A^d(x_i,\hatt\theta)\bigr\}	
	  \bigl\{\psi(x_i,\hatt\theta)\delta_i
			-A^d(x_i,\hatt\theta)\bigr\}^\tr. \cr} \eqno(2.7)$$
Here $\hatt E(t)=\int_0^t\{Y(s)/n\}\psi(s,\hatt\theta)\{\dd N(s)/Y(s)
-\alpha(s,\hatt\theta)\,\dd s\}$ and 
$A^d(t,\theta)=\int_0^t\psi_\theta\alpha_\theta\,\dd s$ is the 
derivative w.r.t.~$\theta$ of $A(t,\theta)=\int_0^t\alpha_\theta\,\dd s$. 
It takes some algebraic skill to show that these are equivalent 
expressions. 
The third formula is computationally more convenient and 
also emerges naturally from the discussion of influence functions 
in the next section. The important statistical consistency property
is however most easily proved using the first formula.

This implies, for an example, that 
the ellipso\"\i d 
$$\bigl\{\theta\colon(\theta-\hatt\theta)^\tr\hatt J\hatt K^{-1}\hatt J
	(\theta-\hatt\theta)\le \gamma_{p,.90}/n\big\}$$
defines an asymptotically correct and model-robust 90\% confidence region
for the most fitting parameter $\theta_0$, when $\gamma_{p,.90}$
is the upper 10\% point of the $\chi^2_p$ distribution.

\smallskip
{\csc Example 2.1.} Study once more the exponential model where 
$\alpha(s,\theta)=\theta$. 
The maximum likelihood estimator is 
$\hatt\theta=N(T)/\int_0^TY(s)\,\dd s=\sum_{i=1}^n\delta_i/\sum_{i=1}^n x_i$.
It converges to the most appropriate parameter value 
$\theta_0=\int_0^Ty(s)\alpha(s)\,\dd s/\int_0^Ty(s)\,\dd s$, i.e.~a $y$-weighted
average of the true hazard rate, by an application of the theorem. 
Furthermore, the second term of the $J$ expression vanishes, and 
$$J={1\over \theta_0^2}\int_0^Ty\alpha\,\dd s, \quad
  K={1\over \theta_0^2}\int_0^Ty\alpha\,\dd s
	+{2\over \theta_0}\int_0^T\int_0^ty(s)\{\alpha(s)-\theta_0\}\,\dd s\,\dd t,$$
with accompanying estimates 
$\hatt J=\{N(T)/n\}/\hatt\theta^2$,
$\hatt K={1\over n}\sum_{i=1}^n(\delta_i/\hatt\theta-x_i)^2$, 
cf.~(2.6) and (2.7). The asymptotic variance of 
$\sqrt{n}(\hatt\theta-\theta_0)$ is estimated by respectively 
$${\hatt\theta^2 \over N(T)/n} \quad {\rm or} \quad 
	{\hatt\theta^4\over \{N(T)/n\}^2} {1\over n}\sum_{i=1}^n
	\Bigl({\delta_i\over \hatt\theta}-x_i\Bigr)^2,$$
under and outside model circumstances.
Note that these expressions reduce to those 
of Example 1.1 when there is no censoring. \firkant
\eject 

\bigskip
\centerline{\bf 3. Influence functions}

\medskip 
This section studies influence functions for estimator functionals
in the presence of censoring, 
and some of their uses are indicated.

The influence function of an estimator is an infinite population concept.
Consider for concreteness the non-censored situation of Section 1 first,
where data come from $F$. 
Assume that an estimator $\hatt\theta$ can be expressed as 
$S(\hatt F)$, where $\hatt F$ is the empirical distribution.
Its target value is $\theta_0=S(F)$. 
The {\it influence function} $I(F,x)$ for such a functional
is the derivative of $S(F_\eps)=S((1-\eps)F+\eps I_x)$ at $\eps=0$,
writing $I_x$ to denote unit point mass at $x$. 
The ordinary maximum likelihood
estimator is for example $\hatt\theta={\rm ml}(\hatt F)$, 
where ${\rm ml}(F)$ is the maximiser of 
$\int\log f_\theta(x)\,\dd F(x)$.
One can demonstrate that 
$$I(F,x)=\lim_{\eps\rightarrow0}
	\eps^{-1}\bigl\{{\rm ml}(F_\eps)-{\rm ml}(F)\}
	=J(F,{\rm ml}(F))^{-1}
		{\dell\log f(x,{\rm ml}(F))\over \dell\theta}, \eqno(3.1)$$
cf.~(1.2) and (1.3) and the linearisation arguments 
that lead to asymptotic distributions. --- Influence functions 
are useful for several purposes. 
It can indicate sensitivity against possible outliers;
it provides a tool with which to find the limit distribution 
of estimators; 
data-based {\it empirical influence functions} 
can be constructed and used to assess 
the influence of individual data points; 
it can sometimes be used to construct new estimators with
specific desiderata; 
and empirical and theoretical influence functions enter 
naturally in studies of the bootstrap and other resampling procedures.  
General references include 
Efron (1982), Reid (1983), 
and Hampel, Ronchetti, Rousseeuw, and Stahel (1986). 

A natural task is now to explore influence functions for 
estimators in the random censorship model of Section 2. 
Reid (1981) and Reid, Cr\'epeau, and Knafl (1985) have also
studied influence functions with censored data, but the present
situation is not covered by their work. 
Let us redescribe the problem 
in a way suiting the task. 
We will limit discussion to the maximum likelihood method. 
The model has been described by saying that partially observed 
$(X_i^0,c_i)$ pairs come from $F\times G$. 
Let $H=H_{F,G}$ be the inherited distribution for data pairs 
$(X_i,\delta_i)=(\min\{X_i^0,c_i\},I\{X_i^0\le c_i\})$ in
$[0,\infty)\times\{0,1\}$. $H$ has subdistribution functions
$H^0(t)={\rm Pr}\{X_i\le t,\delta_i=0\}$ and  
$H^1(t)={\rm Pr}\{X_i\le t,\delta_i=1\}$. 
The data collection can be represented 
by the $N$ and $Y$ processes of (2.1), or equivalently by
the proportion at risk process $\hatt y(s)=Y(s)/n$ with limit 
$y(s)=F[s,\infty)G[s,\infty)$, and the Nelson--Aalen estimator
$\hatt A(t)=\int_0^t\dd N(s)/Y(s)$ with limit $A(t)=\int_0^t\alpha(s)\,\dd s$. 
The $\hatt\theta$ estimator solves
$\int_0^T\psi(s,\theta)\hatt y(s)\{d\hatt A(s)-\alpha(s,\theta)\,\dd s\}=0$ 
and converges to $\theta_0$, the solution of
$\int_0^T\psi(s,\theta)y(s)\{dA(s)-\alpha(s,\theta)\,\dd s\}=0$.
We may view $\theta_0$ as defined by the pair $(F,G)$, or by $(A,y)$,
or by $H=(H^0,H^1)$. 
Observe that $A$ and $y$ can be recovered from $H$, by 
$$\eqalign{
y(s)&={\rm Pr}\{X_i\ge s\}=(H^0+H^1)[s,\infty), \cr
y(s)dA(s)&=dH^1(s)={\rm Pr}\{X_i\in[s,s+\dd s],\delta_i=1\}. \cr}\eqno(3.2)$$ 
In particular, the maximum likelihood method can be viewed as a 
functional ${\rm ml}(H)$ on the space of $H=(H^0,H^1)$ distributions, 
and $\hatt\theta={\rm ml}(\hatt H)$,
where $\hatt H$ is the empirical distribution of 
data pairs $(x_1,\delta_1),\ldots,(x_n,\delta_n)$.

One might consider several influence measures, corresponding
to altering different aspects of the model. 
One can consider variations in ${\rm ml}(A,y)$ 
when $F$ is replaced by $F_\eps=(1-\eps)F+\eps I_x$, 
when $G$ is replaced by $G_\eps=(1-\eps)G+\eps I_c$, 
or both, or replacing $(F,G)$ by $(1-\eps)F\times G+\eps I_{(x,c)}$.
These would give different generalisations 
$I(F,G,x,c)$ of (3.1).
The way data are captured suggests however that we should 
consider local variation of $H$ in the direction of a given
point $(x,\delta)$ in $[0,\infty)\times\{0,1\}$. 

\smallskip
{\csc Theorem 3.1.} {{\sl 
Let $\theta_0={\rm ml}(H)$ for some $H$ under consideration. 
Under the regularity conditions of Theorem 2.1 
the maximum likelihood estimator has influence function
$$\eqalign{
I\bigl(H,(x,\delta)\bigr)&=\lim_{\eps\rightarrow0}
	\eps^{-1}\bigl\{{\rm ml}((1-\eps)H
	    +\eps I_{(x,\delta)})-{\rm ml}(H)\bigr\} \cr
	&=J(H,\theta_0)^{-1}\Bigl[\psi(x,\theta_0)I\{\delta=1\}
	    -{\dell\over \dell\theta}
		\int_0^x\alpha(s,\theta_0)\,\dd s\Bigr] \cr
	&=J(H,\theta_0)^{-1}\int_0^T\psi(s,\theta_0)
		\{\dd N_{x,\delta}(s)
			-Y_{x,\delta}(s)\alpha(s,\theta_0)\,\dd s\}. \cr}$$
Here $J(H,\theta_0)$ is $J(\alpha,y,\theta_0)$ from Section 2, 
and $N_{x,\delta}(t)=I\{x\le t,\delta=1\}$
and $Y_{x,\delta}(s)=I\{x\ge s\}$ are 
counting process and at risk process for the single pair $(x,\delta)$. }}

\smallskip
{\csc Proof:} Write $H_\eps=(1-\eps)H+\eps I_{(x,\delta)}$.
This $H_\eps$ gives rise to $y_\eps$ and $A_\eps$ as follows,
using (3.2):
$$\eqalign{
y_\eps(s)&=(1-\eps)y(s)+\eps\,I\{x\ge s\}, \cr
y_\eps(s)dA_\eps(s)&=(1-\eps)y(s)\,dA(s)
		+\eps\,I\{x\in[s,s+\dd s],\delta=1\}. \cr}$$
We are to find $\theta_\eps={\rm ml}(H_\eps)$, the solution of
$$u_\eps(\theta)=\int_0^T\psi_\theta(s)
	y_\eps(s)\{dA_\eps(s)-\alpha_\theta(s)\,\dd s\}=0.$$
This can be done by 
carrying out a first order Taylor expansion analysis. The result is
$\theta_\eps-\theta_0\doteq\{-{\dell u\over \dell\theta}\}_0^{-1}
\{{\dell u\over \dell\eps}\}_0^{\phantom1}\,\eps$,
where the partial derivatives of $u_\eps(\theta)$ are evaluated at
$\eps=0$ and $\theta=\theta_0$. Some analysis demonstrates that 
$\eps^{-1}(\theta_\eps-\theta_0)$ tends to the limit given in the theorem.
When evaluating ${\dell \over \dell\eps}u_\eps(\theta)$ 
it is crucial to note that $A_\eps$ has a point mass of 
${\rm size}\doteq\eps I\{\delta=1\}/y(x)$ at $x$. 
See also Section 4 for a refinement. \firkant

\smallskip
The result of the theorem generalises (3.1),
since $\log f(x,\theta)=\log\alpha(x,\theta)-A(x,\theta)$ with
derivative $\psi(x,\theta)-A^d(x,\theta)$, and $\delta=1$ in
the non-censored case. 

The result of Theorem 3.1 is also suggested by the proof of Theorem 2.1,
where we in effect showed
$$\hatt\theta-\theta_0={\rm ml}(\hatt H)-{\rm ml}(H)
	\doteq_d J(H,\theta_0)^{-1}{1\over n}\sum_{i=1}^n\int_0^T
	\psi(s,\theta_0)\{\dd N_i(s)-Y_i(s)\alpha(s,\theta_0)\,\dd s\}, \eqno(3.3)$$
writing $N_i$ and $Y_i$ for the counting process and at risk process
of individual no.~$i$. Theorem 2.1 could alternatively have been derived
after Theorem 3.1 using general asymptotic theory of estimators with
influence functions, see e.g.~Reid (1983), Gill (1989),
and the present Section 4. 

Measures of influence for the individual data pairs can be proposed. 
Let $\hatt H_{(i)}$ be the empirical distribution when
$(x_i,\delta_i)$ is deleted from the data set. Then
$$\hatt\theta={\rm ml}\bigl((1-\onebyn)\hatt H_{(i)}
	+\onebyn I_{(x_i,\delta_i)}\bigr)
	 \doteq{\rm ml}(\hatt H_{(i)})+\onebyn 
		I\bigl(\hatt H_{(i)},(x_i,\delta_i)\bigr),$$
which invites using a cross validation type influence measure
$I\bigl(\hatt H_{(i)},(x_i,\delta_i)\bigr)
\doteq n(\hatt\theta-\hatt\theta_{(i)})$
for the $i$'th data pair, where $\hatt\theta_{(i)}$ is computed 
leaving this pair out. It is somewhat simpler to use the approximation
$$\hatt I_i=I\bigl(\hatt H,(x_i,\delta_i)\bigr)
	=J(\hatt H,\hatt\theta)^{-1}\int_0^T\psi(s,\hatt\theta)
	\{\dd N_i(s)-Y_i(s)\alpha(s,\hatt\theta)\,\dd s\}
	=\hatt J^{-1}\hatt L_i \eqno(3.4)$$
instead. Note that $J(\hatt H,\hatt\theta)$ simply is the $\hatt J$
of (2.6), that $\sum_{i=1}^n\hatt I_i=0$, and that
$${1\over n}\sum_{i=1}^n\hatt I_i\hatt I_i^\tr
	=\hatt J^{-1}\Bigl\{{1\over n}\sum_{i=1}^n\hatt L_i\hatt L_i^\tr\Bigr\}
	 \hatt J^{-1}
	=\hatt J^{-1}\hatt K\hatt J^{-1}=\hatt\Sigma, \eqno(3.5)$$
the estimated asymptotic covariance matrix for 
$\sqrt{n}(\hatt\theta-\theta_0)$, cf.~some algebraic manipulations
summed up in (2.7). 

We propose using the $\hatt I_i$'s as a data-analytic tool, to screen
data for possible outliers and to identify data pairs with 
possibly unduly influence. 
A further suggestion is to ``sphere'' them, computing 
$\hatt\Sigma^{-1/2}\hatt I_i
=\hatt J^{1/2}\hatt K^{-1/2}\hatt J^{-1/2}\hatt L_i$ instead. 
These have mean zero and covariance matrix the identity, 
which should make outliers more easily detectable. 

\smallskip
{\csc Remark.} 
Note that we end up with the model-robust
covariance estimator since Theorem 3.1 was derived under the agnostic
point of view. The influence function under model conditions is
similar but with a simpler $J^{-1}$ matrix, see Theorem 2.1. 
As an example, suppose $F_\theta(t)=1-\exp(-t^\theta)$ 
is the Weibull distribution (with a single parameter). 
Then the estimated influence function is
$$\hatt I(x,\delta)=\tilda J^{-1}\bigl\{(1+\log x^{\hatt\theta})\delta
	-x^{\hatt\theta}\log x^{\hatt\theta}\bigr\}/\hatt\theta,$$
where $\tilda J$ is $J(H(.,\hatt\theta),\hatt\theta)$ 
in the model-based case and 
$J(\hatt H,\hatt\theta)
={1\over n}\sum_{i=1}^n\{\delta_i
+x_i^{\hatt\theta}(\log x_i^{\hatt\theta})^2\}/\hatt\theta^2$
in the model-agnostic case. 	
These are different. In the uncensored $[0,\infty)$ case
the first number is simply $1.3504^2/\hatt\theta^2$
[from $1+\Gamma''(2)=(1-\gamma)^2+\pi^2/6=1.3504^2$]. \firkant

\bigskip
\centerline{\bf 4. Model-based and model-robust bootstrapping}

\medskip 
This section briefly studies the large sample behaviour of
some natural bootstrapping schemes. The aim is to use the available
data to come up with simulated versions $\hatt\theta^*$ of 
the maximum likelihood estimator $\hatt\theta$
in such a way that important quantities related to the (partially
unknown) distribution of $\hatt\theta$ can be estimated 
from the empirical distribution of $\hatt\theta^*$.
If interest focusses on some real-valued $\mu=\mu(\theta)$,
then the discussion below applies to $\hatt\mu=\mu(\hatt\theta)$
and $\hatt\mu^*=\mu(\hatt\theta^*)$ instead. 

\smallskip
{\sl 4A. Preliminaries: the maximum likelihood functional.} 
Recall from Section 3 that 
the maximum likelihood procedure can be seen as a functional
operating on distributions $H=(H^0,H^1)$ for $(X,\delta)$.
The estimator aims at ${\rm ml}(H)$, 
the maximiser of $\int_0^Ty(\log\alpha_\theta dA-\alpha_\theta\,\dd s)$,
or, equivalently, the solution of
$\phi(H,\theta)=0$, where
$$\eqalign{
\phi(H,\theta)&=\int_0^Ty(s)\psi(s,\theta)\{dA(s)-\alpha(s,\theta)\,\dd s\} \cr
	&=\int_0^T\psi(s,\theta)\bigl\{dH^1(s)
	 -(H^0+H^1)[s,\infty)\alpha(s,\theta)\,\dd s\bigr\}, \cr}\eqno(4.1)$$
utilising the (3.2) correspondence between $(A,y)$ and $H$
(and we could think of ${\rm ml}(H)$ as ${\rm ml}(A,y)$ instead).   
The nonparametric estimate 
$\hatt H$ for $H$ is the 
empirical distribution of the data pairs $(x_i,\delta_i)$. 
There is a small class of parametric counterparts $H(.,\hatt\theta)$ 
that corresponds to using 
$A(t,\hatt\theta)=\int_0^t\alpha(s,\hatt\theta)\,\dd s$ for $A$ and 
any consistent estimate $\tilda y(t)$ for $y(t)$, for example 
$\hatt y(t)=\exp\{-A(t,\hatt\theta)\}\hatt G[t,\infty)$, 
employing the Kaplan--Meier estimate $\hatt G$ for $G$. 
Observe that both ${\rm ml}(\hatt H)$ and ${\rm ml}(H(.,\hatt\theta))$
indeed are equal to $\hatt\theta$. 

We shall establish that the ml functional is sufficiently smooth,
in a precise sense, and shall have occasion to use this to
rigorously justify that various natural bootstrapping 
schemes actually work. For a pair of distributions $H=(H^0,H^1)$ and
$H_0=(H_0^0,H_0^1)$ for $(X,\delta)$, consider the supremum type norm
$$\eqalign{
\|H-H_0\|^2&=\|H_1^1-H_0^1\|^2+\|H_1^0-H_0^0\|^2 \cr
	   &=\sup_{0\le t\le T}|H_1^1(t)-H_0^1(t)|^2
	     +\sup_{0\le t\le T}|H_1^0(t)-H_0^0(t)|^2. \cr}$$

\smallskip
{\csc Lemma.} 
{{\sl The {\rm ml} functional is locally Lipschitz differentiable 
w.r.t.~the norm $\|H-H_0\|$, under the conditions underlying 
Theorems 2.1 and 3.1. In other words 
$${\rm ml}(H)-{\rm ml}(H_0)=\int_{[0,\infty)\times\{0,1\}} 
	I(H_0,(x,\delta))\,d(H-H_0)(x,\delta)+r(H_0,H),$$
where $r(H_0,H)=O(\|H-H_0\|^2)$ as this distance tends to zero. }}

\smallskip
{\csc Proof:}
Single out some $H_0$ and write $\theta_0={\rm ml}(H_0)$ in what follows.
Consider
$$B(H,(x,\delta))=\int_0^T\psi(s,{\rm ml}(H))
	\bigl\{\dd N_{x,\delta}(s)
		-Y_{x,\delta}(s)\alpha(s,{\rm ml}(H))\,\dd s\bigr\},$$
so that the influence function $I(H,(x,\delta))$ of Theorem 3.1 can be 
written $J(H,{\rm ml}(H))^{-1}\allowbreak B(H,(x,\delta))$. 
Note that $B(.,.)$ acts 
as a functional derivative of $\phi(H,\theta)$ w.r.t.~$H$ in that
$$\phi(H,\theta_0)-\phi(H_0,\theta_0)
	=\int B(H_0,(x,\delta))\,d(H-H_0)(x,\delta)$$
(even without a remainder term). Write for convenience 
$D\phi(H,\theta)$ for the $p\times p$ matrix of (ordinary) 
partial derivatives of $\phi(H,\theta)$ w.r.t.~$\theta$. 
Note that $D\phi(H_0,\theta_0)$ is nothing but the $-J(H_0,\theta_0)$ 
matrix involved in Theorems 2.1 and 3.1. 

We have accordingly derivatives of $\phi(H,\theta)$ in both directions,
and can try Taylor expansion. Assume that 
$$\phi(H,\theta)=\phi(H_0,\theta_0)
	+D\phi(H_0,\theta_0)(\theta-\theta_0)
	+\int B(H_0,(x,\delta))\,d(H-H_0)(x,\delta)
			+r_0(H,\theta) \eqno(4.2)$$
for suitable remainder term $r_0(H,\theta)$. 
Then solving $\phi(H,\theta)=0$ to find ${\rm ml}(H)$ gives
$$\eqalign{
{\rm ml}(H)-{\rm ml}(H_0)&=-D\phi(H_0,\theta_0)^{-1}
	\Bigl[\int B(H_0,(x,\delta))\,d(H-H_0)(x,\delta)
			+r_0(H,{\rm ml}(H)) \Bigr] \cr
&=\int I(H_0,(x,\delta))\,d(H-H_0)(x,\delta)
	+J(H_0,\theta_0)^{-1}r_0(H,{\rm ml}(H)), \cr} $$
and the lemma is proved provided we can show 
$r_0(H,{\rm ml}(H))=O(\|H-H_0\|^2)$. For this is suffices to prove
that $r_0(H,\theta)=O(\|H-H_0\|\,\|\theta-\theta_0\|)$ in (4.2), 
in conjunction with ${\rm ml}(H)-{\rm ml}(H_0)=O(\|H-H_0\|)$. 
But 
$$\eqalign{
r_0(H,\theta)&=\phi(H,\theta)-\phi(H,\theta_0)
		-D\phi(H_0,\theta_0)(\theta-\theta_0) \cr
	&=\bigl[D\phi(H,\theta_0)+O(\|\theta-\theta_0\|)
		-D\phi(H_0,\theta_0)\bigr]\,(\theta-\theta_0) \cr
	&=O(\|H-H_0\|\,\|\theta-\theta_0\|), \cr} $$
using regularity conditions about third order partial 
derivatives etcetera. \firkant  

\smallskip
Suppose $\tilda H$ is some estimate of $H$, 
and let $\tilda H^*$ be the empirical distribution of
data pairs $(x_i^*,\delta_i^*)$ obtained via some scheme or other. Then 
$$\eqalign{
\tilda\theta^*-\tilda\theta
&={\rm ml}(\tilda H^*)-{\rm ml}(\tilda H) \cr
&={1\over n}\sum_{i=1}^nI(\tilda H,(x_i^*,\delta_i^*))
		+r(\tilda H,\tilda H^*) \cr
&=J(\tilda H,\tilda\theta)^{-1}{1\over n}\sum_{i=1}^n
	\int_0^T\psi(s,\tilda\theta)\{\dd N_i^*(s)
		-Y_i^*(s)\alpha(s,\tilda\theta)\,\dd s\} 
		+r(\tilda H,\tilda H^*), \cr} \eqno(4.3)$$
where $N_i^*(t)=I\{x_i^*\le t,\delta_i^*=1\}$ 
and $Y_i^*(t)=I\{x_i^*\ge t\}$ are associated with data pair
$(x_i^*,\delta_i^*)$, cf.~Theorem 3.1. 
To arrive safely at an a.s.~limit distribution result 
for $\sqrt{n}(\tilda\theta^*-\tilda\theta)$ a necessity is
a.s.~convergence to 0 of $\sqrt{n}\,r(\tilda H,\tilda H^*)$.
This follows if $\tilda H^*$ is close enough to $\tilda H$ 
(a statistical question) and ${\rm ml}(.)$ is smooth enough 
(a function space calculus question).
The latter point is dealt with in the lemma. 
Regarding the first point, note that 
if $\tilda H^*$ is the empirical distribution of data from 
$\tilda H$, then $\|\tilda H^*-\tilda H\|=O(\{n^{-1}\log\log n\}^{1/2})$
with probability 1
by well-known fluctuation estimates in the Glivenko--Cantelli theorem, 
from which it follows that 
$\sqrt{n}\|\tilda H^*-\tilda H\|^2=O(n^{-1/2}\log\log n)$ a.s. 
This is also true when $\tilda H$ is non-continuous,
and when $\tilda H=\tilda H_n$ itself is random and converges to 
some fixed $H$, i.e.~$\sqrt{n}\|\tilda H^*_n-\tilda H_n\|^2$ 
is still $O(n^{-1/2}\log\log n)$ a.s.~when $\tilda H^*_n$ is 
the empirical distribution of data from $\tilda H_n$.
See Shao (1989) for similar remarks. 

\smallskip
{\sl 4B. Parametric bootstrapping.}
Simulate pseudo-data $(X_1^*,\delta_1^*),\ldots,(X_n^*,\delta_n^*)$
from the parametrically estimated model. 
In other words, simulate $X_i^{0*}$ from the distribution
with hazard rate $\alpha(.,\hatt\theta)$ and $c_i^*$ from $\hatt G$,
independently, and form $X_i^*=\min\{X_i^{0*},c_i^*\}$, 
$\delta_i^*=I\{X_i^{0*}\le c_i^*\}$. 
(This is actually semi-parametric bootstrapping.) 
Compute $\hatt\theta^*$ from this pseudo-data set, i.e.~from the
empirical distribution $H(.,\hatt\theta)^*$, say, of the $n$ pseudo-pairs.  
Then from (4.3), letting 
$\dd M_i^*(s)=\dd N_i^*(s)-Y_i^*(s)\alpha(s,\hatt\theta)\,\dd s$,  
$$\eqalign{
\sqrt{n}(\hatt\theta_{\rm pb}^*-\hatt\theta)
&=\sqrt{n}\{{\rm ml}(H(.,\hatt\theta)^*)-{\rm ml}(H(.,\hatt\theta))\} \cr
&=J(H(.,\hatt\theta),\hatt\theta)^{-1}
	 {1\over \sqrt{n}}\sum_{i=1}^n\int_0^T\psi(s,\hatt\theta)
			\,\dd M_i^*(s) 
  		+\sqrt{n}\,r(H(.,\hatt\theta),H(.,\hatt\theta)^*). \cr}$$
This can be used to prove
$$\sqrt{n}(\hatt\theta^*_{\rm pb}-\hatt\theta)
	\rightarrow_d N_p\{0,J(H(.,\theta_0),\theta_0)^{-1}\} 
		{\rm \ a.s.} \eqno(4.4)$$
The notation emphasises that there is convergence in distribution with
probability 1, i.e.~the data-conditional distribution 
converges to the right limit for almost all sequences 
of outcomes $(X_i,\delta_i)$.  
Note that the $J$ matrix obtained here is of the  
`under true model' type, and is therefore 
simpler than in the general case
described in Theorem 2.1; in fact 
$$J(H(.,\theta_0),\theta_0)
	=\int_0^Ty(s)\psi(s,\theta_0)\psi(s,\theta_0)^\tr
		\alpha(s,\theta_0)\,\dd s.$$

The first technical point to observe when proving (4.4) 
is that the $M_i^*$'s become orthogonal martingales 
in the conditional framework given data, 
with variance processes $Y_i^*(s)\alpha(s,\hatt\theta)\,\dd s$, 
and that the proof of Theorem 2.1
works in this framework, with $\alpha(s)=\alpha(s,\hatt\theta)$ 
as the underlying true model. 
See Akritas (1988) for somewhat similar arguments carefully spelled out 
in a somewhat similar situation. 
The second point is that the remainder term goes a.s.~to zero,
actually as $O(n^{-1/2}\log\log n)$ by the lemma and the 
remark ending 4A. 

Sometimes $c_i$'s are known, in which case it is natural to just 
put $c_i^*=c_i$ in the bootstrapping scheme above,
or perhaps more information is otherwise 
available about the distribution $G$. 
Suppose $c_i^*$ is drawn from $G_i$ instead of the sometimes coarse 
Kaplan--Meier estimate $\hatt G$. 
The limit distribution argument above rests crucially on 
convergence of $n^{-1/2}\sum_{i=1}^n\int_0^T\psi(s,\hatt\theta)\,\dd M_i^*(s)$.
This is a martingale with variance equal to the mean value 
\line{of $n^{-1}\sum_{i=1}^n\int_0^T\psi(s,\hatt\theta)\psi(s,\hatt\theta)^\tr
Y_i^*(s)\alpha(s,\hatt\theta)\,\dd s$, which is 
$\int_0^T\psi(s,\hatt\theta)\psi(s,\hatt\theta)^\tr
\tilda y(s)\alpha(s,\hatt\theta)\,\dd s$,} 
where $\tilda y(s)=\exp\{-A(s,\hatt\theta)\}\tilda G[s,\infty)$ 
and $\tilda G[s,\infty)=n^{-1}\sum_{i=1}^nG_i[s,\infty)$. 
If only $\tilda G(.)$ tends in probability to the true $G(.)$ 
then martingale limit methods of Helland (1982) can be called upon 
to show that (4.4) holds again. This takes in particular care
of the situation with known $c_i$'s. One has the same (first order)
limit distribution as with $\hatt G$ but presumably 
less sampling variability for fixed $n$.  

\smallskip
{\sl 4C. Nonparametric bootstrapping.}
This time draw $X_i^{0*}$ from the nonparametric 
Kap\-lan--Meier estimate $\hatt F$ instead, 
in tandem with an independent $c_i^*$ from $\hatt G$, as above. 
This happens to be equivalent to drawing $(X_i^*,\delta_i^*)$ pairs
independently from $\hatt H$, as explained in Efron (1981). 
Somewhat more elaborate arguments are needed in this case. 
Let $\dd M_i^*(s)=\dd N_i^*(s)-Y_i^*(s)\,d\hatt A(s)$. 
The $M_i^*$'s become orthogonal martingales in the data-conditional 
framework, with variance process $Y_i^*(s)d\hatt A(s)\{1-d\hatt A(s)\}$. 
From (4.3) 
$$\eqalign{
\sqrt{n}(\hatt\theta^*_{\rm nb}-\hatt\theta)
&=\sqrt{n}\{{\rm ml}(\hatt H^*)-{\rm ml}(\hatt H)\} \cr
&=J(\hatt H,\hatt\theta)^{-1}{1\over \sqrt{n}}
	\sum_{i=1}^n\int_0^T\psi(s,\hatt\theta)
	\bigl[\dd M_i^*(s)+Y_i^*(s)\{d\hatt A(s)
		-\alpha(s,\hatt\theta)\,\dd s\}\bigr] \cr 
	 & \qquad +\sqrt{n}\,r(\hatt H,\hatt H^*). \cr}$$		
The remainder term again goes a.s.~to zero by the efforts of 4A,
and $J(\hatt H,\hatt\theta)$, which is $\hatt J$ of (2.6), 
is strongly consistent for $J=J(H,\theta_0)$ under the present conditions.
The middle term can be written 
$$\int_0^T\psi(s,\hatt\theta)\Bigl[\dd M^*(s)/\sqrt{n}
	+\sqrt{n}\{Y^*(s)/n-\hatt y(s)\}\,
		\{d\hatt A(s)-\alpha(s,\hatt\theta)\,\dd s\}\Bigr]$$
and resembles an expression used in the proof of Theorem 2.1.
This proof can in fact be copied and used in the present problem
with suitable delicate alterations, to show that the middle term
tends in distribution a.s.~to $N_p\{0,K(H,\theta_0)\}$, 
where the $K$ matrix is as in Theorem 2.1. 
The details require some modest machinery for discrete time
martingales, as in Helland (1982), and can be taken care 
of by means similar to those in the Appendix of Hjort (1985b).   
The end result is 
$$\sqrt{n}(\hatt\theta^*_{\rm nb}-\hatt\theta)
	\rightarrow_d N_p\{0,J^{-1}KJ^{-1}\} 
	{\rm \ a.s.} \eqno(4.5)$$ 

\smallskip
{\sl 4D. Discussion.} 
The consequences of (4.4) and (4.5) are more or less as 
for the classical non-censored case, discussed briefly after (1.7).
The nonparametric bootstrap always works correctly, 
in the first order large sample sense, as a consequence of (4.5) 
and Theorem 2.1. The parametric bootstrap creates the correct amount
of variability only if the model itself is correct. Otherwise 
either under- or overestimation could result. 
(4.4) is statistically meaningful even when the model is wrong, 
in that it tells about the estimation uncertainty in a situation with
data from a correct model at the least false $\theta_0$. 
If the model does happen to be adequate, then 
both $\hatt\theta_{\rm nb}^*$ and $\hatt\theta_{\rm pb}^*$ have the
same limit distributions, but the nonparametric one will usually
have larger sampling variability. This is for example clear when
one writes down the necessary expressions in the 
situation with censored data from an exponential distribution. 

There are other bootstrapping schemes. We noted that all sensible
ways of drawing $c_i^*$'s in the parametric case gives the same
large sample behaviour for $\hatt\theta_{\rm pb}^*$. This is not
quite the case for $\hatt\theta^*_{\rm nb}$. If one uses 
the empirical distribution $\tilda G$ in the case of known $c_i$'s,
then the nonparametric scheme with $X_i^{0*}$'s from $\hatt F$
is first of all not equivalent to drawing pairs $(X_i^*,\delta_i^*)$'s 
from $\hatt H$ anymore, and secondly the limit distribution
of $\sqrt{n}(\hatt\theta^*_{\rm nb}-\hatt\theta)$ exists but is
slightly different from that of $\sqrt{n}(\hatt\theta-\theta_0)$. 

Our justification proof for the bootstrap schemes used 
local Lipschitz differentiability of the ml functional.
Results (4.4) and (4.5) could have been reached in other ways 
as well. Rather general function space methods in Gill (1989) 
and Cs\"org\H o and Mason (1989) could be used, but would give somewhat
weaker results, without the extra bonus of speed of convergence
which our Lipschitz method gives. 
On the other hands the methods used by these authors would give 
results even without the almost sure convergence details that partly 
underlie our proof, and this is relevant in more complex 
counting process models where perhaps only weak consistency 
can be proved for $\hatt\theta$.  
It is also worth pointing out
that the technical matters were helped by the assumed 
finiteness of the observation interval $[0,T]$. With likelihoods
on the full halfline $[0,\infty)$ the ml functional would not be 
quite Lipschitz differentiable, and there would also have been
difficulties with applying the implicit function theorem, 
when solving for $\theta$ in $\phi(H,\theta)=0$, if one were to 
use Gill's machinery. 

We note finally that our asymptotic results also justify the
use of so-called bootstrap-$t$ procedures, 
to `first order'. These are sometimes more precise 
on the `second order' level; see comments in 7D. 

\bigskip
\centerline{\bf 5. Other estimation methods}

\medskip
We have concentrated on the maximum likelihood estimator $\hatt\theta$
in previous sections. Hjort (1986a, Section 3) proved 
that several of the familiar asymptotic
optimality properties enjoyed by this method in classical situations 
carry over to the present censored data framework. These
properties have however as basic assumption that the 
parametric model is indeed correct.
There is therefore still interest in studying other estimation schemes,
that perhaps might be somewhat less inefficient than $\hatt\theta$ under
the ideal model's home turf conditions  
but that for example could have better robustness
properties outside model conditions. 
This section briefly discusses some possibilities. 

\smallskip
{\sl 5A. Bayes estimators.} 
If $\pi(\theta)\,d\theta$ is a prior
density for $\theta$ then the Bayes estimator is 
$\hatt\theta_B=E\{\theta|{\rm data}\}=
\int\theta L_n(\theta)\pi(\theta)\,d\theta/
\int L_n(\theta)\pi(\theta)\,d\theta$.
But as far as first order asymptotic behaviour is concerned 
such estimators are equivalent to 
the maximum likelihood solution, i.e.~$\sqrt{n}(\hatt\theta_B-\hatt\theta)$
goes to zero in probability, even outside model conditions, according
to Hjort (1986a, Section 2). 

\smallskip
{\sl 5B. M-type estimators.} 
We saw in Example 2.1 that the maximum likelihood solution in the
constant hazard rate model tends to 
$\theta_0=\int_0^Ty\alpha\,\dd s/\int_0^Ty\,\dd s$,
a weighted average of the true hazard rate over the observation interval.
As a consequence small $s$-values are given much more weight 
than larger $s$-values. 
Perhaps more disturbing is the fact that
the somewhat problem-irrelevant censoring distribution $G$ is involved
in $\theta_0$, through $y(s)=F[s,\infty)G[s,\infty)$. 
This is a general feature of the maximum likelihood approach, see (2.3).  
One could argue that the most fitting constant hazard rate
should be $\theta_1=\int_0^T\alpha\,\dd s/\int_0^T \dd s$ instead,
or at least that it should be freed of its dependence upon $G$.

This corresponds
to a different weighting of the log-likelihood. Consider in general terms
the {\it weighted likelihood} 
$$WL_n(\theta)=\exp\Bigl[\int_0^TW_n(s)\bigl\{\log\alpha_\theta(s)\,\dd N(s)
	-Y(s)\alpha_\theta(s)\,\dd s\bigr\}\Bigr], \eqno(5.1)$$
where $W_n(.)$ is a weight function tending in probability to some $w(.)$,
and where the notation is as in Section 2. The corresponding  
maximum weighted likelihood estimator $\hatt\theta_w$ maximises this
function, and also solves 
$\int_0^TW_n\psi_\theta\{\dd N-Y\alpha_\theta\,\dd s\}=0$.
An alternative term suggested by an analogy to the 
non-censored i.i.d.~situation is {\it M-estimators}.
These and more general estimators were 
introduced in the counting process model 
context in Hjort (1985, Section 4), where  
limiting distribution results were given,
but only under model conditions. 
Similarly motivated estimators in a different
stochastic process framework have been studied in S\o rensen (1990). 

It is now possible to go through the arguments of Section 2 and 3 
and apply them to M-estimators. Under appropriate and mild regularity
conditions, which include $W_n(s)\rightarrow_p w(s)$, 
it holds that $n^{-1}\log WL_n(\theta)$ tends to
$\int_0^Twy(\alpha\log\alpha_\theta-\alpha_\theta)\,\dd s$, 
that $\hatt\theta_w$ is consistent for the (new)
least false parameter $\theta_{0,w}$ that minimises the differently
weighted distance measure
$$d_w[\alpha,\alpha_\theta]=\int_0^Twy\bigl\{
	\alpha(\log\alpha-\log\alpha_\theta)
	-(\alpha-\alpha_\theta)\bigr\}\,\dd s, \eqno(5.2)$$ 
cf.~(2.3), in particular each $M$-estimator is consistent at the model, 
and that 
$$\sqrt{n}(\hatt\theta_w-\theta_{0,w})\rightarrow_d J_w^{-1}N_p\{0,K_w\}
	=N_p\{0,J_w^{-1}K_wJ_w^{-1}\}, \eqno(5.3)$$
in which 
$$J_w=\int_0^Twy\bigl[\psi(.,\theta_{0,w})\psi(.,\theta_{0,w})^\tr
	 \alpha(.,\theta_{0,w})
	 -D\psi(.,\theta_{0,w})
	 \{\alpha-\alpha(.,\theta_{0,w})\}\bigr]\,\dd s,$$
$$K_w={\rm VAR}\int_0^Tw(s)\psi(s,\theta_{0,w})\bigl[dV(s)
	+Z(s)\{\alpha(s)-\alpha(s,\theta_{0,w})\}\,\dd s\bigr],$$
cf.~(2.4). We point out that the weight function $W_n(s)$ is allowed to
be random here, it can for example be previsible (its value at 
time $s$ is known at time $s-$), or of the form $G_n(s,\hatt\theta)$,
where $G_n(s,\theta_{0,w})$ is previsible and converges to $w(s,\theta_{0,w})$
in probability. (Such a function's value at time $s$ is {\it not} 
known at time $s-$, since it employs $\hatt\theta$, which requires all the
$[0,T]$-data to be computed.) 

This apparatus can now be used to construct a 
{\it modified maximum likelihood estimator} 
$\hatt\theta_m$ that avoids being dependent upon 
the censoring distribution $G$. The point is to use 
$W_n(s)=\hatt G[s,\infty)^{-1}$,
where $\hatt G[s,\infty)=\prod_{u\le s}\{1-\dd N_c(u)/Y(u)\}$
is the Kaplan--Meier
estimator based on the observed censoring times. 
The accompanying distance measure for $\hatt\theta_m$ is (5.2) above with 
$y(s)w(s)=y(s)G[s,\infty)^{-1}=F[s,\infty)=\exp\{-A(s)\}$,
and is perhaps an even more appropriate generalisation of 
Kullback--Leibler's information distance than (2.3), see Remark 7A.
The modified $\hatt\theta_m$ is consistent for $\theta_{0,m}$,
for example, $\theta_{0,m}=\int_0^Te^{-A}\alpha\,\dd s/\int_0^Te^{-A}\,\dd s$ 
in the exponential model. This points out anew that different estimators
might converge to different least false values when the model is incorrect;
$\hatt\theta_m$ aims here at a value more tied to the 
`inverse expected time to failure' interpretation of $\theta$ than to the 
`constant hazard rate' interpretation. 

Another interesting choice is 
$W_n(s)=\hatt y(s)^{-1}=\hatt F[s,\infty)^{-1}\hatt G[s,\infty)^{-1}$.
It converges to $y(s)^{-1}$ and has the effect of freeing the estimator
from its dependence on $y(.)$, i.e.~from favouring portions of $[0,T]$
with large $y$ over portions with small $y$. In the exponential case
this modificator estimates $\theta_{0,w}=\int_0^T\alpha(s)\,\dd s/T$, 
the neutrally weighted hazard rate. 

Using the modified estimator entails
a loss in efficiency at the model, 
as $J_w^{-1}K_wJ_w^{-1}$ is a larger matrix than $J^{-1}$. 
As an example, study the exponential model, 
suppose that $\alpha(s)=\theta_0$ prevails,  
and assume that the censoring distribution is $G(t)=1-\exp(-g\theta_0)$,
which corresponds to an expected frequency $1/(g+1)$ of $(x_i,\delta_i)$
pairs where $x^0_i$ is truly observed.
The maximum likelihood estimator $\hatt\theta$ 
and the two modificators $\hatt\theta_{m1}$ and $\hatt\theta_{m2}$
mentioned above all take the form
$\int_0^TW_n \dd N/\int_0^TW_nY\,\dd s$, using respectively
$W_n(s)=1$, $W_n(s)=\hatt G[s,\infty)^{-1}$, and 
$W_n(s)=\hatt y(s)^{-1}$. All three are consistent for $\theta_0$
(since the model is in command), and their asymptotic variances 
can be shown to be respectively 
$${1\over n}{\theta_0^2\over 1-\eps}, \quad
	{1\over n}{\theta_0^2\over 1-g}{1-\eps^{1-g}\over (1-\eps)^2}, \quad
	{1\over n}{\theta_0^2\over 1+g}
			{(1/\eps)^{1+g}-1\over (\log1/\eps)^2},$$
in which ${\rm Pr}\{X^0\le T\}=1-\exp(-\theta_0T)=1-\eps$.
The third estimator is too defensive in its avoidance of the model,
and is much worse than the two others for most combinations of $g$ and $\eps$.
The second estimator does not lose much efficiency for values of
$g$ that signal low or moderate amounts of censoring, say $g\le \half$. 
The efficiency loss becomes significant in cases with more than a moderate
amount of censoring.     

The influence function of an M-estimator can also be found, 
using arguments presented in Section 3. 
With notation as there it becomes 
$$I\bigl(H,(x,\delta)\bigr)=J_w^{-1}\int_0^Tw(s)\psi(s,\theta_{0,w})
	\{\dd N_0(s)-Y_0(s)\alpha(s,\theta_{0,w})\,\dd s\}, \eqno(5.4)$$
and an estimator for it can easily be constructed,
along with empirical influence measures of the type 
$\hatt I_i=I\bigl(\hatt H,(x_i,\delta_i)\bigr)$.  
If the maximum likelihood method looks non-robust, in that
the influence function given in Theorem 3.1 is sensitive to large
values of $x$, then a more robust estimator can be constructed 
by using an appropriate deflating $w$-function.

\smallskip
{\sl 5C. Dynamic likelihood and smoothing.} 
A choice of $W_n$ different in spirit from those considered above 
is $W_n(s)=w(s)=I\{s\in B\}$, for a suitable subinterval $B$ of $[0,T]$.
The resulting estimator $\hatt\theta_B$ uses only data 
for individuals who are at risk at the beginning of $B$ 
and information about what happens to them during $B$, 
and aims at a locally most appropriate
$\theta_{0,B}$, the parameter value that minimises 
$d_B[\alpha,\alpha_\theta]$, say, which is as in (2.3) but integrated 
only over $B$. Such estimates could be computed for different subintervals
and compared, for example for model checking purposes. 
See Hjort (1990) for such and for much more 
general model checking procedures. 

A similar but more ambitious idea, both statistically and computationally,
is to use a local $B(s)=(s-\half h,s+\half h]$ around each given $s$, 
to compute an estimate $\hatt\theta(s)=\hatt\theta_{B(s)}$ 
based only on local data.
This corresponds to a dynamic or local likelihood approach, 
and is somewhat similar in motivation
to work by Hastie and Tibshirani (1987).
Choosing once again the constant hazard rate model as an example, 
$\hatt\theta(s)=N\{B(s)\}/\int_{B(s)}Y(s)\,\dd s$ 
becomes the dynamic hazard rate estimate at $s$. 
This is similar to but not the same as kernel smoothing 
of the Nelson--Aalen estimator $\int_0^.\dd N/Y$, 
which is Ramlau-Hansen's (1983) way 
of nonparametrically estimating a hazard rate. 
We can also pass to general kernel function smoothing, 
and for each $s$ maximise
$$\int_0^TK(s-u)\bigl\{\log\alpha_\theta(u)\,\dd N(u)
	-Y(u)\alpha_\theta(u)\,du\bigr\}$$
to obtain the local or dynamic or smoothing $\hatt\theta(s)$, where
$K$ is symmetric with maximum at zero. 
We view these methods as semiparametric approaches to the estimation
of a parametric model. 
A more complete development of this and other
semiparametric estimation methods is in Hjort (1991b). 
%
Observe that the local likelihood 
method also can be used to construct a 
``dynamic semiparametric density estimator'' via 
$f_\theta(t)=\alpha_\theta(t)\exp\{-A_\theta(t\}$, and of course
works in cases without censoring as well. 
A dynamic estimator of the normal density can for example 
easily be constructed, of the form
$\hatt f(t)=N\{\hatt\mu(t),\hatt\sigma^2(t)\}$, where
$\hatt\mu(t)$ and $\hatt\sigma(t)$ are obtained locally. 
These matters will be pursued elsewhere. 

\bigskip
\centerline{\bf 6. Regression models for hazard rates}

\medskip
So far we have considered lifetimes to have been 
drawn from a homogeneous population. Statistically more challenging
and important problems arise when the individuals under study also
have covariate measurements that may influence the lifetime distribution.
In this section two regression models for hazard rates are studied,
the traditional semiparametric Cox model with unspecified baseline hazard rate 
and the fully parametric Cox model with parametric baseline hazard rate. 
Once more the questions to be discussed include 
behaviour of the maximum 
likelihood estimators outside the narrow model assumptions, 
agnostic estimation of the covariance matrix, 
and influence measures. 

The data set is $(x_1,\delta_1,z_1),\ldots,(x_n,\delta_n,z_n)$, where
$x_i$ and $\delta_i$ are as in previous sections and $z_i$ is a $q$-dimensional
covariate measurement vector for individual no.~$i$. 
The hazard rate for this individual is $\alpha_i(s)=\alpha(s|z_i)$.
The Cox model postulates that
$$\alpha_i(s)=\alpha(s)\,\exp(\beta^\tr z_i)
	=\alpha(s)\,\exp(\beta_1z_{i,1}+\cdots\beta_qz_{i,q}), 
				\quad i=1,\ldots,n, \eqno(6.1)$$
where $\alpha(.)$ is an unspecified hazard rate 
and $\beta$ is a vector of coefficients. These are traditionally
estimated by maximum partial likelihood, see Gill (1984) for a 
good account of the theory. However, the behaviour of the estimates
outside the narrow proportional hazards assumption seems not to have 
been studied in the literature, except for Hjort (1986a), where the
point limit $\beta_0$ of the estimates is identified outside model
conditions. In 6B below also the limit distribution is found, and a
consistent estimate is provided for the covariance matrix. 

The success of Cox regression analysis has perhaps had the unintended 
side effect that practitioners too seldomly invest efforts in studying the
baseline hazard $\alpha(.)$. A parametric version, say
$$\alpha_i(s)=\alpha(s,\theta)\,\exp(\beta^\tr z_i), 
				\quad i=1,\ldots,n, \eqno(6.2)$$
for some $p$-dimensional $\theta$, if found to be adequate, 
would lead to more precise estimation of survival probabilities and
related quantities and concurrently contribute to a better understanding
of the survival phenomenon under study. 
This is the model studied in 6A below. 
References where such models
have been used, with $\alpha(s,\theta)$ corresponding to the exponential,
Weibull, log-normal distribution, or to piece-wise constant hazards,
can be found in Kalbfleisch and Prentice (1980, Chapter 3) and Borgan (1984). 
Hjort (1990a) provides goodness of fit tests for models of type (6.2).

\smallskip
{\csc Remark.}
For ease of exposition we shall assume throughout this section that 
data are {\it i.i.d.~realisations} 
of a triple $(X,\Delta,Z)$, with appropriate distribution $H$ in
$[0,\infty)\times\{0,1\}\times{\cal R}^q$, 
and that $(X,\Delta)$ come from life time $X^0$ and censoring time $C$ 
in the manner described in Section 2. 
The sequence $(N_i,Y_i,M_i)$ of individual analogues to (2.1), 
that is $\dd N_i(s)=I\{X_i\in[s,s+\dd s],\Delta_i=1\}$, $Y_i(s)=I\{X_i\ge s\}$,
and $M_i(t)=N_i(t)-\int_0^tY_i(s)\alpha_i(s)\,\dd s$, also become i.i.d.
The i.i.d.-assumption is not crucial; 
most of the reasoning and results below continue to be valid
for example in a setting with non-random censoring times and covariates,
with suitable modifications. \firkant

\smallskip
{\sl 6A. Parametric Cox regression.} 
The model postulates (6.2). Assume that the true state of affairs is 
of the form $\alpha_i(s)=\alpha(s|z_i)=\alpha_0(s)h_0(z_i)$
for some $\alpha_0(.)$ and some $h_0(.)$;  
this is the hazard rate that would have been seen if a large data set
were collected from individuals with the same covariate vector $z_i$. 
The maximum likelihood estimators $\hatt\theta$, $\hatt\beta$ maximise
$${1\over n}\log L_n=
{1\over n}\sum_{i=1}^n\int_0^T\Bigl[\{\log\alpha(s,\theta)+\beta^\tr z_i\}
	\,\dd N_i(s)-Y_i(s)\alpha(s,\theta)\exp(\beta^\tr z_i)\,\dd s\Bigr].$$
They also solve $U_n(\theta,\beta)=0$, where $U_n$ has components
$$\eqalign{
U_n^{(1)}(\theta,\beta)
	&={1\over n}\sum_{i=1}^n\int_0^T
	\psi_\theta\{\dd N_i-Y_i\alpha_\theta\exp(\beta^\tr z_i)\,\dd s\}
	=\int_0^T\psi_\theta\{\dd G_n^{(0)}-Q_n^{(0)}\alpha_\theta\,\dd s\}, \cr
U_n^{(2)}(\theta,\beta)
	&={1\over n}\sum_{i=1}^n\int_0^T
	z_i\{\dd N_i-Y_i\alpha_\theta\exp(\beta^\tr z_i)\,\dd s\}
	=\int_0^T\{\dd G_n^{(1)}-Q_n^{(1)}\alpha_\theta\,\dd s\}. \cr}$$
A certain amount of extra notation is necessary here. We use
$$\eqalign{
R_n^{(0)}(s)&={1\over n}\sum_{i=1}^nY_i(s)h_0(z_i)\rightarrow_p
	r^{(0)}(s)=EI\{X\ge s\}h_0(Z), \cr
R_n^{(1)}(s)&={1\over n}\sum_{i=1}^nY_i(s)z_ih_0(z_i)\rightarrow_p
	r^{(1)}(s)=EI\{X\ge s\}Zh_0(Z), \cr}$$
$$\eqalign{
\dd G_n^{(0)}(s)
&={1\over n}\sum_{i=1}^n \dd N_i(s)\rightarrow_p
	 \dd G^{(0)}(s)=EI\{X\in[s,s+\dd s],\Delta=1\}=r^{(0)}(s)\alpha_0(s)\,\dd s, \cr
\dd G_n^{(1)}(s)
&={1\over n}\sum_{i=1}^nz_i\,\dd N_i(s)\rightarrow_p
	\dd G^{(1)}(s)=EZI\{X\in[s,s+\dd s],\Delta=1\}
		=r^{(1)}(s)\alpha_0(s)\,\dd s, \cr}$$
$$\eqalign{
Q_n^{(0)}(s,\beta)&={1\over n}\sum_{i=1}^nY_i(s)\exp(\beta^\tr z_i)
	\rightarrow_p
	q^{(0)}(s,\beta)=EI\{X\ge s\}\exp(\beta^\tr Z), \cr
Q_n^{(1)}(s,\beta)&={1\over n}\sum_{i=1}^nY_i(s)z_i\exp(\beta^\tr z_i)
	\rightarrow_p
	q^{(1)}(s,\beta)=EI\{X\ge s\}Z\exp(\beta^\tr Z), \cr
Q_n^{(2)}(s,\beta)&={1\over n}\sum_{i=1}^nY_i(s)z_iz_i^\tr \exp(\beta^\tr z_i)
	\rightarrow_p
	q^{(2)}(s,\beta)=EI\{X\ge s\}ZZ^\tr\exp(\beta^\tr Z). \cr}$$
If the model is perfect, then $h_0(z)=\exp(\beta_0^\tr z)$ 
for some $\beta_0$ and $R_n^{(0)}(s)=Q_n^{(0)}(s,\beta_0)$, 
$r^{(0)}(s)=q^{(0)}(s,\beta_0)$, etc.

To study the behaviour of the estimators, observe that the components
of $U_n(\theta,\beta)$ have limits 
$$\eqalign{
u^{(1)}(\theta,\beta)&=\int_0^T\psi_\theta\{\dd G^{(0)}
	 -q^{(0)}(.,\beta)\alpha_\theta\,\dd s\}
	=\int_0^T\psi_\theta\{r^{(0)}\alpha_0
	 -q^{(0)}(.,\beta)\alpha_\theta\}\,\dd s, \cr
u^{(2)}(\theta,\beta)&=\int_0^T\{\dd G^{(1)}
	 -q^{(1)}(.,\beta)\alpha_\theta\,\dd s\}
	=\int_0^T\{r^{(1)}\alpha_0
	 -q^{(1)}(.,\beta)\alpha_\theta\}\,\dd s. \cr} \eqno(6.3)$$
These functions determine the limit $(\theta_0,\beta_0)$ 
of $(\hatt\theta,\hatt\beta)$, see the theorem below. 
Taking second partial derivatives of $n^{-1}\log L_n(\theta,\beta)$ gives
a matrix $I_n(\theta,\beta)$, and an expression for 
its limit in probability can be found. Let $J$ be the limit of 
$-I_n(\theta_0,\beta_0)$. It has blocks
$$J_{11}=\int_0^Tq^{(0)}(.,\beta_0)\psi(.,\theta_0)\psi(.,\theta_0)^\tr
	\alpha(.,\theta_0)\,\dd s-\int_0^TD\psi(.,\theta_0)
	\{r^{(0)}\alpha_0-q^{(0)}(.,\beta_0)\alpha(.,\theta_0)\}\,\dd s,$$
$$J_{12}=\int_0^T\psi(.,\theta_0)q^{(1)}(.,\beta_0)^\tr
	\alpha(.,\theta_0)\,\dd s, \quad 
  J_{22}=\int_0^Tq^{(2)}(.,\beta_0) 
	\alpha(.,\theta_0)\,\dd s.$$
Let finally 
$$K={\rm VAR}\left[\matrix{
	\int_0^T\psi(s,\theta_0)\{\dd N_i(s)-Y_i(s)\alpha(s,\theta_0)
		\exp(\beta_0^\tr Z_i)\,\dd s\} \cr
	\int_0^TZ_i\{\dd N_i(s)-Y_i(s)\alpha(s,\theta_0)
		\exp(\beta_0^\tr Z_i)\,\dd s\} \cr}\right].$$
A somewhat complicated explicit expression can be given for $K$,
as in the proof of Theorem 2.1, this time involving $\alpha_0$ and $h_0$, 
but we will be content with this
description and the consistent estimator below.
The following result generalises a theorem of Borgan (1984)
to outside-the-model conditions. 

\smallskip
{\csc Theorem 6.1.} {{\sl 
Assume that the equations $u^{(1)}(\theta,\beta)=0$,
$u^{(2)}(\theta,\beta)=0$ have a unique solution $(\theta_0,\beta_0)$. 
Suppose further that the regularity conditions on $\alpha(s,\theta)$ 
stated in Theorem 2.1 hold, that $J$ is positive definite,
that the covariates $z_i$ are uniformly bounded as $n$ grows, 
and that $h_0(z)$ is bounded away from zero and infinity in this
bounded domain.
Then the maximum likelihood estimators 
$(\hatt\theta,\hatt\beta)$ are consistent
for the least false parameter values $(\theta_0,\beta_0)$. 
These also minimise the distance measure 
$d[\alpha_0(.)h_0(.),\alpha_\theta(.)\exp(\beta^\tr.)]$ given in (6.6) below.
Furthermore,
$$\left[\matrix{
	\sqrt{n}(\hatt\theta-\theta_0) \cr
	\sqrt{n}(\hatt\beta-\beta_0) \cr}\right]
	\rightarrow_d J^{-1}N_{p+q}\{0,K\}=N_{p+q}\{0,J^{-1}KJ^{-1}\},$$
where $J$ and $K$ are given above. A consistent estimator for $J$ is $\hatt J$,
with blocks
$$\hatt J_{11}=\int_0^TQ_n^{(0)}(.,\hatt\beta)
	\psi(.,\hatt\theta)\psi(.,\hatt\theta)^\tr
	\alpha(.,\hatt\theta)\,\dd s-\int_0^TD\psi(.,\hatt\theta)
	\{\dd G_n^{(0)}(s)-Q_n^{(0)}(.,\hatt\beta)\alpha(.,\hatt\theta)\,\dd s\},$$
$$\hatt J_{12}=\int_0^T\psi(.,\hatt\theta)Q_n^{(1)}(.,\hatt\beta)^\tr
	\alpha(.,\hatt\theta)\,\dd s, \quad 
  \hatt J_{22}=\int_0^TQ_n^{(2)}(.,\hatt\beta) 
	\alpha(.,\hatt\theta)\,\dd s.$$
Finally  
$$\hatt K={1\over n}\sum_{i=1}^n\hatt L_i\hatt L_i^\tr, \quad
	\hatt L_i=\left[\matrix{
	\psi(x_i,\hatt\theta)\delta_i
		-\exp(\hatt\beta^\tr z_i)A^d(x_i,\hatt\theta) \cr
	z_i\{\delta_i-\exp(\hatt\beta^\tr z_i)A(x_i,\hatt\theta)\} \cr}\right]$$
is a consistent estimator for $K$. }}

\smallskip
We note that the regularity conditions can be weakened, along
the lines of Borgan (1984, Section 6), but that those given here
should be satisfied in most practical applications. 
Note also that uniqueness of the root of $u(\theta,\beta)=0$,
or of the minimiser of the (6.6) distance, 
follows if the log-likelihood function is concave.

\smallskip 
{\csc Proof:}
The consistency part can essentially be handled using methods of 
Hjort (1986a, Theorem 2.3). The asymptotic normality part is similar to 
the proof of Theorem 2.1, again using the Taylor expansion argument
(1.2). One has to employ the martingales 
$M_i(t)=N_i(t)-\int_0^tY_i(s)\alpha_0(s)h_0(z_i)\,\dd s$
where Borgan (1984) was allowed by the model to use 
$N_i(t)-\int_0^tY_i(s)\alpha(s,\theta_0)\exp(\beta_0^\tr z_i)\,\dd s$,
and take the additional variability into account.   
The variables whose covariance matrix defines $K$ above split into 
a martingale term and an additional more complicated term that
comes from incorrectness of the model; only under model circumstances
does the second term vanish and $K$ become equal to $J$. 
Consistency of $\hatt J$ and $\hatt K$ can be established using 
martingale inequalities and uniform convergence in probability arguments
that for example can be gleaned from Hjort (1990a, Section 2). 
Let's leave it at that. \firkant

\smallskip
Measures of influence become even more important in the presence of covariates.
Let $H$ be the distribution of $(X,\Delta,Z)$, 
and let $(x,\delta,z)$ be fixed. Then $H_\eps=(1-\eps)H+\eps I_{(x,\delta,z)}$
represents a small perturbation of $H$ in direction $(x,\delta,z)$,
and the least false $(\theta_\eps,\beta_\eps)$ determined by
$H_\eps$ can be studied. It is by the theorem the solution to 
$u_\eps^{(1)}(\theta,\beta)=0$, $u_\eps^{(2)}(\theta,\beta)=0$, where
$u_\eps$ is as in (6.3), but with
$$\eqalign{
\dd G^{(j)}_\eps(s)&=(1-\eps)\dd G^{(j)}(s)
	+\eps z^j\,I\{x\in[s,s+\dd s],\delta=1\}, \cr
q^{(j)}_\eps(s,\beta)&=(1-\eps)q^{(j)}(s,\beta)
		+\eps z^j\,I\{x\ge s\}\exp(\beta^\tr z), \cr} \eqno(6.4)$$
for $j$ equal to 0 and 1. Note that $G_\eps^{(0)}$ and $G_\eps^{(1)}$
have positive point masses at $x$ if $\delta=1$. 
Analysis as in the simpler case covered by Theorem 3.1 
gives at the end of the night the influence function 
$$\eqalign{
I\bigl(H,(x,\delta,z)\bigr)&=\lim_{\eps\rightarrow0}
	 \left[\matrix{\{\theta_0(H_\eps)-\theta_0(H)\}/\eps \cr
		      \{\beta_0(H_\eps)-\beta_0(H)\}/\eps \cr}\right] \cr
	&=J^{-1}\left[\matrix{
	 \psi(x,\theta_0)\delta
		-\exp(\beta_0^\tr z)A^d(x,\theta_0) \cr
	 z\{\delta-\exp(\beta_0^\tr z)A(x_i,\theta_0)\} \cr}\right] \cr
	&=J^{-1}\left[\matrix{
	 \int_0^T\psi(s,\theta_0)\{\dd N_0(s) 
	  -Y_0(s)\alpha(s,\theta_0)\exp(\beta_0^\tr z)\,\dd s\} \cr
	 \int_0^Tz\{\dd N_0(s)
	  -Y_0(s)\alpha(s,\theta_0)\exp(\beta_0^\tr z)\,\dd s\} \cr}\right], \cr}$$
in which $N_0$ and $Y_0$ are counting process and at risk process for 
$(x,\delta)$.
Natural diagnostic measures for influence are 
$$\hatt I_i=I\bigl(\hatt H,(x_i,\delta_i,z_i)\bigr)=\hatt J^{-1}\hatt L_i
	=\hatt J^{-1}\left[\matrix{
	\int_0^T\psi(s,\hatt\theta)\{\dd \dd N_i(s)
		-Y_i(s)\alpha(s,\hatt\theta)\exp(\hatt\beta^\tr z_i)\,\dd s\} \cr
	\int_0^Tz_i\{\dd N_i(s)
		-Y_i(s)\alpha(s,\hatt\theta)\exp(\hatt\beta^\tr z_i)\,\dd s\} \cr}
							\right], \eqno(6.5)$$
where an alternative expression for $\hatt L_i$ is given in the theorem, 
and $\hatt H$ is the empirical distribution of the 
$n$ triples $(x_i,\delta_i,z_i)$. 
It is also an approximation to
the crossvalidated $I\bigl(\hatt H_{(i)},(x_i,\delta_i,z_i)\bigr)$ and to 
$\bigl(n(\hatt\theta-\hatt\theta_{(i)}),n(\hatt\beta-\hatt\beta_{(i)})\bigr)$,
see Section 3. A further important property of these empirical influence 
measures is that their empirical covariance matrix becomes
$\hatt\Sigma=\hatt J^{-1}\hatt K\hatt J^{-1}$, as in (3.5). 
We propose computing the sphered influence measures  
$\hatt\Sigma^{-1/2}\hatt I_i$, which have mean zero and 
empirical covariance matrix the identity in dimension $p+q$, 
to screen data for outliers and for 
individual data triples with particular influence. 

Let us end this subsection with exhibiting the distance measure
between hazard rates with respect to which $(\theta_0,\beta_0)$
chosen by the maximum likelihood procedure is least false,
cf.~the first part of the Introduction. We reach slightly more general
insight by writing $\alpha(s|z)=\alpha_\theta(s)h_\beta(z)$ for
the parametric model, instead of the special case (6.2), 
and $\alpha_0(s)h_0(z)$ for the true model. 
Under these circumstances one can show that
$${1\over n}\log L_n(\theta,\beta)
	 \rightarrow_p
	 \int_0^T\bigl\{r^{(0)}(s)\log\alpha_\theta(s)\,\alpha_0(s) 
	 +r^{(1)}(s,\beta)\alpha_0(s)
	 -q^{(0)}(s,\beta)\alpha_\theta(s)\bigr\}\,\dd s,$$
where the functions entering the integrand are given below, and also
expressed as integrals over
the covariate space ${\cal Z}$ with respect to the covariate
distribution $D(dz)$ for $Z$:
$$\eqalign{
r^{(0)}(s)&=EI\{X\ge s\}h_0(Z)=\int_{\cal Z}y(s|z)h_0(z)\,D(dz), \cr
r^{(1)}(s,\beta)&=EI\{X\ge s\}h_0(Z)\log h_\beta(Z)
		=\int_{\cal Z}y(s|z)h_0(z)\log h_\beta(z)\,D(dz), \cr
q^{(0)}(s,\beta)&=EI\{X\ge s\}h_\beta(Z)
		=\int_{\cal Z}y(s|z)h_\beta(z)\,D(dz). \cr}$$
These equations also feature the $z$-dependent 
$y(s|z)={\rm Pr}\{X\ge s|z\}$. Consider the $z$-dependent 
hazard distance from $\alpha_0(.)h_0(z)$ to $\alpha_\theta(.)h_\beta(z)$,
as measured by the already encountered distance measure (2.3), that is
$$\eqalign{
d_z[\alpha_0(.)h_0(z),\alpha_\theta(.)h_\beta(z)]
	 =\int_0^Ty(s|z)\Bigl[&\alpha_0(s)h_0(z)
	 \log{\alpha_0(s)h_0(z)\over \alpha_\theta(s)h_\beta(z)} \cr
	 &-\{\alpha_0(s)h_0(z)
		-\alpha_\theta(s)h_\beta(z)\}\Bigr]\,\dd s. \cr}$$
It is now a matter of careful checking to see that maximising the limit of
$n^{-1}\log L_n(\theta,\beta)$ is the same as minimising the 
$z$-weighted distance function
$$d[\alpha_0\,h_0,\alpha_\theta\,h_\beta]
	=\int_{\cal Z}d_z[\alpha_0(.)h_0(z),\alpha_\theta(.)h_\beta(z)]
		\,D(dz). \eqno(6.6)$$

\smallskip
{\sl 6B. Semiparametric Cox regression.}
The model postulates (6.1), where $\alpha(.)$ is left unspecified.
Let us, conservatively and counterbalancedly, 
assume only that $\alpha_i(s)=\alpha(s|z_i)=\alpha(s)h_0(z_i)$ for
some $\alpha(.)$ and some $h_0(.)$. The Cox estimator maximises the
partial log likelihood 
$$\log L_n(\beta)=\sum_{i=1}^n\int_0^T\Bigl[\beta^\tr z_i
	-\log\bigl\{\sum_{j=1}^nY_j(s)\exp(\beta^\tr z_j)\bigr\}\Bigr]\,\dd N_i(s),$$
see for example Gill (1984). It is also a root of 
$$U_n(\beta)={1\over n}\sum_{i=1}^n\int_0^T\Big\{
	z_i-{Q_n^{(1)}(s,\beta)\over Q_n^{(0)}(s,\beta)}\Bigr\}\,\dd N_i(s)
	=\int_0^T\bigl\{\dd G_n^{(1)}(s)-E_n(s,\beta)\,\dd G_n^{(0)}(s)\bigr\},$$
where notation is as in 6A and $E_n=Q_n^{(1)}/Q_n^{(0)}$, with limit
$e(s,\beta)=q^{(1)}(s,\beta)/q^{(0)}(s,\beta)$. We have
$$U_n(\beta)\rightarrow_pu(\beta)=\int_0^T\bigl\{\dd G^{(1)}(s)
	-e(s,\beta)\,\dd G^{(0)}(s)\bigr\}
	=\int_0^T\bigl\{r^{(1)}(.)-{q^{(1)}(.,\beta)\over q^{(0)}(.,\beta)}
		r^{(0)}(.)\bigr\}\,dA(s). \eqno(6.7)$$
If the model is perfect, then $r^{(0)}=q^{(0)}(.,\beta_0)$ and
$r^{(1)}=q^{(1)}(.,\beta_0)$ for some $\beta_0$, and in particular 
$u(\beta_0)=0$. The consistency part of the theorem below generalises this;
once more there is a least false parameter $\beta_0$ 
even when the (6.1) model is incorrect. 

We shall also need the second order partial derivatives of $n^{-1}\log L_n$,
aiming once more at establishing limit distributions via Taylor
expansion and (1.2). One finds 
$$\eqalign{
-I_n(\beta_0)&=\int_0^T\Bigl\{
	{Q_n^{(2)}(s,\beta_0)\over Q_n^{(0)}(s,\beta_0)}
	-E_n(s,\beta_0)E_n(s,\beta_0)^\tr \Bigr\}\,\dd G_n^{(0)}(s) \cr
	&\rightarrow_p
	J=\int_0^T\Bigl\{
	{q^{(2)}(s,\beta_0)\over q^{(0)}(s,\beta_0)}
	-e(s,\beta_0)e(s,\beta_0)^\tr \Bigr\}\,r^{(0)}(s)\,dA(s). \cr}$$
Observe that the formula usually given in the literature for 
this information matrix has the model-based $q^{(0)}(s,\beta_0)$ 
in lieu of our agnostic $r^{(0)}(s)$. 
Let finally
$$K={\rm VAR}\int_0^T\{Z-e(s,\beta_0)\}\Bigl\{\dd N_0(s)
	-Y_0(s)\exp(\beta_0^\tr Z){r^{(0)}(s)\over q^{(0)}(s,\beta_0)}
		\alpha(s)\,\dd s\Bigr\},$$
where $N_0(t)=I\{X\le t,\Delta=1\}$, $Y_0(s)=I\{X\ge s\}$, and $(X,\Delta,Z)$
has distribution $H$. A long and complicated explicit expression 
can be obtained for $K=K(H)$, but the description here suffices for our
purposes. What is important is knowing the existence of this
matrix and how it enters the limit distribution, 
and having an explicit consistent estimator, which we provide below.

\smallskip
{\csc Theorem 6.2.} {{\sl 
Suppose that the $(X_i,\Delta_i,Z_i)$`s are i.i.d.,
as stated in the remark before 6A, and that 
$Z_i=(Z_{i,1},\ldots,Z_{i,q})$ has a finite moment 
generating function (in particular, uniform boundedness
of the $Z_i$'s suffice). 
Suppose further that the functions $q^{(0)}(s,0)$ 
and $r^{(0)}(s)$ of 6A are bounded away from zero 
and infinity when $0\le s\le T$,
and finally that the $J$ and $K$ matrices 
given above are positive definite. 
Then the Cox estimator $\hatt\beta$ 
is consistent for the least false parameter 
value $\beta_0$ that uniquely solves $u(\beta)=0$. 
This parameter value also minimises the distance function 
$d[h_0(.),\exp(\beta^\tr .)]$ given in (6.9) below. Furthermore
$$\sqrt{n}(\hatt\beta-\beta_0)\rightarrow_dJ^{-1}N_q\{0,K\}
	=N_q\{0,J^{-1}KJ^{-1}\},$$
and
$$\hatt J=\int_0^T\Bigl\{
	{Q_n^{(2)}(s,\hatt\beta)\over Q_n^{(0)}(s,\hatt\beta)}
	-E_n(s,\hatt\beta)E_n(s,\hatt\beta)^\tr \Bigr\}\,\dd G_n^{(0)}(s),$$
$$\hatt K={1\over n}\sum_{i=1}^n\hatt L_i\hatt L_i^\tr , \quad
	\hatt L_i=\int_0^T\{z_i-E_n(s,\hatt\beta)\}\Bigl\{\dd N_i(s)
	-Y_i(s)\exp(\hatt\beta^\tr z_i)
	{\dd G_n^{(0)}(s)\over Q_n^{(0)}(s,\hatt\beta)}\Bigr\}$$
are consistent estimators. }}

\smallskip
{\csc Proof:} Consistency and uniqueness of $\beta_0$ 
follows from the methods and results of 
Hjort (1986a, Theorem 4.1 and its proof).  
Methods provided there are also sufficient to
demonstrate $-I_n(\tilda\beta)\rightarrow_p J$ in the appropriate 
analogue of (1.2). What remains to be shown, therefore, is that 
$\sqrt{n}\,U_n(\beta_0)\rightarrow_d N_q\{0,K\}$. This can be done along the
lines of the proof of Theorem 2.1, though matters become much more 
involved. One establishes that 
$$\sqrt{n}\,U_n(\beta_0)\doteq_d{1\over \sqrt{n}}\sum_{i=1}^n
	\{z_i-E_n(s,\beta_0)\}\,\dd M_i(s)
	+\int_0^T\sqrt{n}\{W_n(s)-w(s)\}\,dA(s),$$
where $M_i(t)=N_i(t)-\int_0^tY_i(s)\alpha(s)h_0(z_i)\,\dd s$ is a martingale 
and where 
$$W_n(s)={1\over n}\sum_{i=1}^nY_i(s)\{z_i-e(s,\beta_0)\}
	\Bigl\{h(z_i)-{r^{(0)}(s)\over q^{(0)}(s,\beta_0)}
		\exp(\beta_0^\tr z_i)\Bigr\}$$
with expected value $w(s)=r^{(1)}(s)-e(s,\beta_0)r^{(0)}(s)$.
Note that our $M_i$ is different from the traditionally employed 
$N_i(t)-\int_0^tY_i(s)\alpha(s)\exp(\beta_0^\tr z_i)\,\dd s$, 
see Andersen and Gill (1982) and Gill (1984).
These authors found themselves in the 
luxurious possession of a perfect model,
in which case the second term above 
vanishes, since $W_n(s)$ then is zero. 
The rest of the proof therefore generalises the model-based proof 
by establishing joint convergence in distribution of 
the martingale $n^{-1/2}\sum_{i=1}^n\int_0^.\{z_i-E_n(s,\beta_0)\}\,\dd M_i(s)$
and $\sqrt{n}\{W_n(s)-w(s)\}$, in the function space 
$D[0,T]\times D[0,T]$, and finally computing the covariance matrix, which
indeed becomes $K$. Inserting consistent estimators for unknown parameters
and functions in this expression gives a consistent estimator $\tilda K$.
A long algebraic exercise reminiscent of the manipulations that led to
the simplified third expression in (2.7) 
shows that $\tilda K=\hatt K$.~\firkant

\smallskip
Let us also provide the influence function for the semiparametric Cox model. 
This can be found by analysis similar to that carried out in 6A
for the parametric Cox model. 
Suppose $\beta_0(H_\eps)$ is the least false parameter vector under
$H_\eps=(1-\eps)H+\eps I_{(x,\delta,z)}$, i.e.~the solution to
$u_\eps(\beta)=0$, where $u_\eps$ is as in (6.7) but with 
appropriate $\dd G_\eps^{(0)}$ and $\dd G_\eps^{(1)}$ instead, as in (6.4). 
One finds that $\{\beta_0(H_\eps)-\beta_0(H)\}/\eps$ tends to
$\{-{\dell u\over \dell\beta}\}_0^{-1}
\{{\dell u\over \dell\eps}\}_0^{\phantom1}$,
where the partial derivatives are evaluated at $\beta=\beta_0$ and
$\eps=0$. The first matrix 
$\{-{\dell u\over \dell\beta}\}_0^{\phantom1}$ is simply $J$. 
Taking the point masses at $x$ for both $G_\eps^{(0)}$
and $G_\eps^{(1)}$ into account one reaches
$$\eqalign{
I\bigl(H,(x,\delta,z)\bigr)&=J^{-1}\Bigl[
	\Bigl\{z-{q^{(1)}(x,\beta_0)\over q^{(0)}(x,\beta_0)}\Bigr\}\delta
	-\int_0^x{\exp(\beta_0^\tr z)\over q^{(0)}(s,\beta_0)}
	\Bigl\{z-{q^{(1)}(s,\beta_0)\over q^{(0)}(s,\beta_0)}\Bigr\}
						\,\dd G^{(0)}(s)\Bigr] \cr 
	&=J^{-1}\int_0^T\Bigl\{z
		-{q^{(1)}(s,\beta_0)\over q^{(0)}(s,\beta_0)}\Bigr\}
	\Bigl\{\dd N_0(s)-Y_0(s)\exp(\beta_0^\tr z)
		{r^{(0)}(s)\over q^{(0)}(s,\beta_0)}
			\,\alpha(s)\,\dd s\Bigr\}; \cr}$$
$N_0$ and $Y_0$ belong once more to the single triple $(x,\delta,z)$.

The discussion ending subsection 6A can now be repeated with small changes.
The empirical influence function is $I(\hatt H,(x,\delta,z))$,
and the natural influence measure for data triple $(x_i,\delta_i,z_i)$
becomes 
$$\hatt I_i=I\bigl(\hatt H,(x_i,\delta_i,z_i)\bigr)
	=\hatt J^{-1}\hatt L_i, \eqno(6.8)$$
where $\hatt L_i$ is given the Theorem 6.2. 
These sum to zero and have empirical covariance matrix equal to the important 
$\hatt\Sigma$, the estimate for the limiting covariance matrix
of $\sqrt{n}(\hatt\beta-\beta_0)$. 
The sphered versions $\hatt\Sigma^{-1/2}\hatt I_i$ have 
the identity matrix as empirical covariance matrix, 
and sore thumbs should stick out.

Reid, Cr\'epeau, and Knafl (1985) also gave an influence function 
for the Cox regression model. They used another method and did not
make it clear that their evaluations in fact were valid also outside
the model conditions. They reached an influence measure in their formula
(2), given in a form very different from ours, but it turns out to be
identical to (6.8). 

Let us finally provide the distance measure under which 
the $\beta_0$ parameter chosen by the Cox method is least false,
in the spirit of the introductory remarks of Section 1. 
Let us be slightly more general and allow $\alpha(s|z)=\alpha(s)h_\beta(z)$
for the model, instead of (6.1), 
and suppose the truth is $\alpha(s)h_0(z)$. Then
${1\over n}\log L_n(\beta)$ can be shown to converge in probability to
$$\lambda(h_0,h_\beta)=\int_0^T\bigl\{r^{(1)}(s,\beta)
	-r^{(0)}(s)\log q^{(0)}(s,\beta)\bigr\}\alpha(s)\,\dd s,$$
using the same notation as in (6.6). One can now show that
the maximum of $\lambda(h_0,g)$ over all $g$ functions is $\lambda(h_0,h_0)$.
[One possibility is to prove it first in the simple case of a finite support 
$\{z_1,\ldots,z_m\}$ for the design variable distribution $D(dz)$ for $Z$,
where the problem becomes one of maximising a given function with respect
to $g(z_1),\ldots,g(z_m)$. Then one can pass to the general case with
appropriate limit arguments.] Hence there is a natural distance measure
with respect to which Cox's maximum partial likelihood estimator 
converges to the least false value:
$$\eqalign{
d&[h_0(.),h_\beta(.)]=\lambda(h_0,h_0)-\lambda(h_0,h_\beta) \cr
	&=\int_0^T\Bigl[EI\{X\ge s\}h_0(Z)\log{h_0(Z)\over h_\beta(Z)}
	 -EI\{X\ge s\}h_0(Z)\log{EI\{X\ge s\}h_0(Z) \over 
		EI\{X\ge s\}h_\beta(Z)} \Bigr]\,dA(s) \cr
	&=\int_{\cal Z}\int_0^Ty(s|z)\Bigl[\log{h_0(z)\over h_\beta(z)} 
	 -\log{EI\{X\ge s\}h_0(z)\over EI\{X\ge s\}h_\beta(z)}\Bigr]
		dA(s)\,h_0(z)D(dz). \cr} \eqno(6.9)$$

\bigskip
\centerline{\bf 7. Discussion and concluding remarks}

\medskip 
In this final section a couple of complementary remarks are offered,
some of which point to further research.

\smallskip
{\sl 7A. Some identities in the absence of censoring.} 
General formulae were derived under
censoring circumstances in Section 2, 
and these should reduce to the more familar ones of Section 1 when
no censoring is present and the observation period is $[0,\infty)$.
Without censoring the $y$ of (2.2) is simply $\exp(-A)$,
writing $A$ and $A_\theta$ for the cumulative hazard rates. 
The identities below are valid in this $y=\exp(-A)$ case.

The new formula for the limit of $n^{-1}\log L_n(\theta)$ is
$\int_0^Ty(\alpha\log\alpha_\theta-\alpha_\theta)\,\dd t$. 
The densities can be written 
$f_\theta=\alpha_\theta\exp(-A_\theta)$ and $f=\alpha\exp(-A)$.
Integration by parts yields
$$\int_0^Ty(\alpha\log\alpha_\theta-\alpha_\theta)\,\dd t
	=\int_0^Tf\log f_\theta\,\dd t-e^{-A(t)}A_\theta(T), \eqno(7.1)$$
and we have $\int_0^\infty f\log f_\theta\,\dd t$ when $T$ grows.
When this identity for $\alpha$ is applied also to $\alpha_\theta$,
we find for the new distance measure (2.3) between hazard rates
$$d[\alpha,\alpha_\theta]=\int_0^Tf\log(f/f_\theta)\,\dd t
	-e^{-A(T)}\{A(T)-A_\theta(T)\}. \eqno(7.2)$$
Accordingly this distance generalises the Kullbak--Leibler information 
distance. 

The new formula for the limit of 
$n^{-1}\dell\log L_n(\theta)/\dell\theta$ is 
$\int_0^Ty\psi_\theta(\alpha-\alpha_\theta)\,\dd t$.
Taking partial derivatives of the first identity gives
$$\int_0^Ty\psi_\theta(\alpha-\alpha_\theta)\,\dd t
	=\int_0^Tf{\dell\log f_\theta\over \dell\theta}\,\dd t
	 -e^{-A(T)}\int_0^T\alpha_\theta\psi_\theta\,\dd t,$$
and we have the appropriate limit when $T$ grows.
Taking second order partial derivatives of the same identity yields
$$\eqalign{
\int_0^Te^{-A}\bigl[\psi_\theta\psi_\theta^\tr\alpha_\theta
	&-D\psi(.,\theta)(\alpha-\alpha_\theta)\bigr]\,\dd t \cr
&=-\int_0^Tf{\dell^2\log f_\theta \over \dell\theta\dell\theta}\,\dd t 
 +e^{-A(T)}\int_0^T\bigl[\psi_\theta\psi_\theta^\tr 
		+D\psi(s,\theta)\bigr]\alpha_\theta\,\dd t. \cr}$$
In particular the $J$ matrix of Section 2 becomes 
$-E_F\dell^2\log f_\theta/\dell\theta\dell\theta$ when $T$ reaches 
infinity.

Consider finally the $K$ matrix. One can show that
$$\eqalign{
K&=\int_0^Te^{-A}\psi_{\theta_0}(\psi_{\theta_0})^\tr \alpha\,\dd t
	+\int_0^T\bigl[E(\psi_{\theta_0})^\tr+\psi_{\theta_0}E'\bigr]
		\alpha_{\theta_0}\,\dd t \cr
	&=\int_0^T\bigl(\psi_{\theta_0}-A^d_{\theta_0}\bigr)
		\bigl(\psi_{\theta_0}
	    	-A^d_{\theta_0}\bigr)' e^{-A}\alpha\,\dd t	
		+e^{-A(T)}A^d_{\theta_0}(T)A^d_{\theta_0}(T)', \cr}$$
where $A^d_{\theta}(t)=\int_0^t\alpha_\theta(s)\psi_\theta(s)\,\dd s$ 
is the derivative of $A_\theta(t)$ w.r.t.~$\theta$. Note that the usual
score function is the derivative of the logarithm of
$f_\theta(t)=\alpha_\theta(t)\exp\{-A_\theta(t)\}$,
that is, $L_\theta(t)=\psi_\theta(t)-A^d_\theta(t)$. 
In the limit as $T$ grows we have 
$K=\int_0^\infty L_\theta(t)L_\theta(t)^\tr\,\dd F(t)$, as we should. 

\smallskip
{\sl 7B. General counting process models.} 
For ease of exposition our basic framework has been that of the
random censorship model. Most of our arguments use martingale 
theory only, however, and go through with minor modifications for
general and multivariate parametric counting process models,
see Andersen and Borgan (1985) and 
Andersen, Borgan, Gill and Keiding (1992) 
for reviews of relevant methods.
One particular detail that does become more difficult is that 
of almost sure convergence of the maximum likelihood estimator.
In the structurally simplest versions of a parametric counting process 
models, as in Borgan (1984) and Hjort (1986a), only convergence 
in probability has been established. This does not affect the theory of
Sections 2, 3, 5, but some small amendments are called for regarding
the equivalent of Section 4 for such general models. 
The principal difference is that results (4.4) and (4.5) for the
bootstrap must be phrased differently;
the bootstrap distributions converge in probability only.
This will follow by applying the apparatus of Section 4 
without Lipschitz diffentiability but with Hadamard differentiability
instead, see Gill (1989, Section 4). 
The methods of Cs\"org\H o and Mason (1989) 
could conceivably also be used.

\smallskip
{\sl 7C. Bootstrapping in regression models for survival data.}
Section 4 treated only homogeneous models. 
Consider for concreteness the parametric Cox model (6.2)
for data $(X_i,\delta_i,z_i)$ with distribution $H$.  
More than simply `model-based' and `model-robust' bootstrapping
schemes can be proposed in such a situation. 
Scheme 1 could be to generate $z_i^*$ from some estimated 
covariance distribution, nonparametric or parametric, 
and then $X_i^{0*}$ from the distribution 
with hazard $\alpha(s,\hatt\theta)\,\exp(\hatt\beta^\tr z_i^*)$ 
along with $c_i^*$ from some suitable $G_i$, for example the
Kaplan--Meier estimate for the censoring distribution.
One might also just keep $z_i^*=z_i$ for individual $i$. 
This scheme gives one way of obtaining $(X_i^*,\delta_i^*,z_i^*)$,
trying to be as faithful to the postulated model as possible.
Scheme 2 could be to resample triplets, i.e.~from the
empirical distribution $\hatt H$. This method ignores all the
finer structure of the model. Scheme 3 could be in the semiparametric
Cox spirit and simulate $X_i^{0*}$ 
from the estimated distribution 
$\hatt F_i(t)=1-\prod_{[0,t]}\{1-d\hatt A(s)\}^{\exp(\hatt\beta^\tr z_i)}$,
cf.~Hjort (1985b, Section 1). Scheme 4 could use a nonparametric
smoother for the relative risk part instead of $\exp(\hatt\beta^\tr z_i)$.
As indicated each of these schemes will have its sub-schemes.

The first order behaviour of all these schemes can be
sorted out with the methods developed in this paper, 
under and outside model conditions. This also goes for 
similar schemes for the semiparametric Cox model.
This careful cataloguing is left for future work.  
Let us merely mention one result, which judicious 
calculations will show: 
All schemes indicated above are first order asymptotically
correct if the (6.2) model is correct, in the sense that
$(\sqrt{n}(\hatt\theta^*-\hatt\theta),
\sqrt{n}(\hatt\beta^*-\hatt\beta))^\tr $ has the same limiting
distribution, with probability 1, as 
$(\sqrt{n}(\hatt\theta-\theta_0),\sqrt{n}(\hatt\beta-\beta_0))^\tr $. 
See the first part of Hjort (1985b) for the kind of 
arguments that would be needed, 
in addition to Sections 3 and 4 of the present paper.  
Scheme 1 would however display smaller sampling variability 
than Scheme 2.

A general point raised by a referee is that the 
$(\delta_i,z_i)$'s in some sense are ancillaries,
so that statistical inference should be carried out 
conditionally on the observed values of these statistics.
And bootstrapping does this. 

\smallskip
{\sl 7D. Finer bootstrap analysis.} 
Our study has been a first order large sample one, regarding
both behaviour of estimates and of bootstrapped versions of them.
One could enter the more difficult world of second order
expansions and second order correct confidence intervals as well.
At least in the random censorship model it should be possible to
show that bootstrapping based on studentised statistics
provide second order correct intervals, that is, 
approximate the distribution of 
$t=\sqrt{n}\{\mu(\hatt\theta)-\mu(\theta_0)\}/\hatt\tau$
with that of 
$t^*=\sqrt{n}\{\mu(\hatt\theta^*)-\mu(\hatt\theta)\}/\hatt\tau^*$,
where $\hatt\tau$ is an estimate of the limiting standard deviation for
$\sqrt{n}\{\mu(\hatt\theta)-\mu(\theta_0)\}$ and $\hatt\tau^*$ its
bootstrap sister. Methods of Hall (1988) are relevant here,
as would second order methods for martingales,
as rudimentarily presented in the Appendix of Hjort (1985b). 
In the latter paper second order correct intervals of 
Efron's ABC variety are constructed for the parameters 
in Cox' regression model.

\smallskip
{\sl 7E. Implications.}
This paper has provided precise large-sample 
results on consequences of
misspecification of the underlying survival data model. 
We have discussed various general implications of these results, 
but have avoided going into detailed analysis 
of concrete examples. Such analysis, like studying 
the actual difference between the true $\alpha(.)$ curve
to the best approximant $\alpha(.,\theta_0)$,
and the resulting errors in covering probabilities of
confidence intervals, for example, 
would be interesting and fruitful, 
for a suitable list of often-used models and 
realistic deviations from them. 
And one could similarly explore
consequences for the $\beta$'s in a parametric Cox model 
with misspecified parametric part $\alpha(s,\theta)$,
as a referee has suggested. 
Of relevance here is also the work on 
moderate misspecification presented in Hjort (1991a). 

\bigskip
{\bf Acknowledgments.} 
The editor and referees gave useful comments and references, 
and I have enjoyed conversations with 
Ishani Manjula Arulchelvam and 
Cristina Maria Trist\~ao Sim\~oes Rocha. 

\bigskip
\font\bf=cmbx9
\font\rm=cmr9
\font\sl=cmsl9 
\centerline{\bf References}

\medskip\parindent0pt
\baselineskip12pt
\parskip2pt
\rm

Akritas, M.G. (1986). Bootstrapping the Kaplan--Meier estimator.
{\sl J.~Amer.~Statist.~Assoc.}~{\bf 81}, 1032--1038.

Andersen, P.K.~and Gill, R.D. (1982).
Cox's regression model for counting processes: a large sample study.
{\sl Ann.~Statist.}~{\bf 10}, 1100--1120.

Andersen, P.K.~and Borgan, \O. (1985).
Counting process models for life history data: A review (with
discussion). {\sl Scand.~J.~Statist.}~{\bf 12}, 97--158.

Andersen, P.K., Borgan, \O., Gill, R., and Keiding, N. (1992).
{\sl Counting Process Models.} Springer.

Barndorff-Nielsen, O.~and S\o rensen, M. (1991).
Asymptotic likelihood theory for stochastic processes.
{\sl Int.~Statist.~Rev.}, to appear. 

Billingsley, P. (1968). {\sl Convergence of Probability Measures.}
Wiley, New York.

Borgan, \O. (1984). Maximum likelihood estimation in parametric counting
process models, with applications to censored failure time data.
{\sl Scand.~J.~Statist.}~{\bf 11}, 1--16. Corrigendum, {\it ibid.}~p.~275.

Chibisov, D.M. (1973).
An asymptotic expansion for a class of estimators containing
maximum likelihood estimators.
{\sl Theory Prob.~Appl.}~{\bf 18}, 295--303. 

Cox, D.R. (1962).
Further results on tests on separate families of hypotheses. 
{\sl J.~Royal Statist.~Soc. B}~{\bf 24}, 406--424.

Cs\"org\H o, S.~and Mason, D.M. (1989).
Bootstrapping empirical processes.
{\sl Ann.~Statist.}~{\bf 17}, 1447--1471. 

Efron, B. (1981). Censored data and the bootstrap.
{\sl J.~Amer.~Statist.~Assoc.}~{\bf 76}, 312--319.

Efron, B. (1982). {\sl The Jackknife, the Bootstrap, and Other Resampling
Plans.} SIAM--NSF, CBMS \#38, Philadelphia.

Fernholz, L.T. (1983). {\sl von Mises calculus for statistical
functionals.} Lecture Notes in Statistics. Springer, New York.

Gill, R.D. (1984).
Understanding Cox's regression model: a martingale approach.
{\sl J.~Amer. Sta\-tist.~Assoc.}~{\bf 79}, 441--447.

Gill, R.D. (1989).
Non- and semi-parametric maximum likelihood estimation and the 
von Mises method (part I, with discussion).
{\sl Scand.~J.~Statist.}~{\bf 16}, 97--128.  

Hall, P. (1988). 
Theoretical discussion of bootstrap confidence
intervals (with discussion).
{\sl Ann. Statist.}~{\bf 16}, 927--985. 

Hampel, F.R., Ronchetti, E.M., Rousseeuw, P.J., and Stahel, W.A. (1986).
{\sl Robust Statistics: The Approach Based on Influence Functions.}
Wiley, Singapore.

Hastie, T.J.~and Tibshirani, R. (1986).
Generalized additive models (with discussion).
{\sl Statist.~Science}~{\bf 1}, 297--318.

Helland, I. (1982). 
Central limit theorems for martingales with 
discrete or continuous time.
{\sl Scand.~J.~Statist.}~{\bf 9}, 79--94.

Hjort, N.L. (1985a). 
Discussion contribution to Andersen and Borgan's
review article. {\sl Scand.~J. Statist.}~{\bf 12}, 141--150.

Hjort, N.L. (1985b). 
Bootstrapping Cox's regression model.
Technical Report NSF--241, Department of Statistics, Stanford University.

Hjort, N.L. (1986a). 
Bayes estimators and asymptotic efficiency
in parametric counting process models. 
{\sl Scand.~J.~Statist.}~{\bf 13}, 63--85.

Hjort, N.L. (1986b).
Discussion contribution to Diaconis and Freedman's
``On the consistency of Bayes estimates''.
{\sl Ann.~Statist.}~{\bf 14}, 49--55.

Hjort, N.L. (1988). 
Discussion contribution to Hinkley's lectures
on bootstrapping techniques. To appear in 
{\sl Scand.~J.~Statist.}

Hjort, N.L. (1990). 
Goodness of fit tests in models for life history
data based on cumulative hazard rates. 
{\sl Ann.~Statist.}~{\bf 18}, 1221--1258.

Hjort, N.L. (1991a).
Estimation in moderately misspecified models.
Statistical Research Report, University of Oslo;
submitted for publication. 

Hjort, N.L. (1991b).
Semiparametric estimation of parametric hazard rates.
Proceedings of the {\sl NATO Advanced Study Workshop
on Survival Analysis and Related Topics},
Columbus, Ohio. 

Huber, P.J. (1967).
The behaviour of maximum likelihood estimators
under nonstandard conditions.
{\sl Proc.~Fifth Berkeley Symp.~on Math.~Statist.~and Probab.},
University of California Press, 221--233. 

Kalbfleisch, J.D.~and Prentice, R.L. (1980).
{\sl The Statistical Analysis of Failure Time Data.}
Wiley, Singapore.

Lehmann, E.L. (1983). {\sl Theory of Point Estimation.}
Wiley, Singapore. 

Lin, D.Y.~and Wei, L.J. (1989).
The robust inference for the Cox 
proportional hazards model. 
{\sl J.~Amer.~Statist.~Assoc.}~{\bf 84}, 1074--1078.

Linhart, H.~and Zucchini, W. (1986).
{\sl Model Selection.} Wiley, Singapore. 

McKeague, I.W. (1984).
Estimation for diffusion processes
under misspecified models. 
{\sl J.~Appl. Probab.}~{\bf 21}, 511--520. 

Ramlau-Hansen, H. (1983). 
Smoothing counting process intensities by means of kernel functions.
{\sl Ann.~Statist.}~{\bf 11}, 453--466.

Reeds, J.A. (1978). 
Jackknifing maximum likelihood estimates.
{\sl Ann.~Statist.}~{\bf 6}, 727--739.

Reid, N. (1981). Influence functions for censored data. 
{\sl Ann.~Statist.}~{\bf 9}, 78--92. 

Reid, N. (1983). Influence functions. 
In {\sl Encyclopedia of Statistical Science} {\bf 4}, eds.~Kotz, Johnson,
Read, 117--119. Wiley, New York.

Reid, N., Cr\'epeau, H., and Knafl, G. (1985).
Influence functions for proportional hazards regression.
{\sl Biometrika} {\bf 72}, 1--9.

Shao, J. (1989). 
Functional calculus and asymptotic theory for statistical analysis.
{\sl Statist. and Probab.~Letters}~{\bf 8}, 397--405.

S\o rensen, M. (1990).
On quasi likelihood for semimartingales.
{\sl Stoch.~Proc.~Appl.}~{\bf 35}, 331--346. 

White, H. (1982).
Maximum likelihood estimation of misspecified models.
{\sl Econometrica} {\bf 50}, 1--26.

\baselineskip12pt
\font\rm=cmr10
\font\sl=cmsl10
\font\bf=cmbx10

\bigskip\bigskip\noindent
{\bf R\'esum\'e.}
Les mod\`eles parametriques habituels pour l'analyse
des don\'ees de survie ont la forme suivante:
on consid\`ere une fonction de risque param\'etrique specifique
$\alpha(s,\theta)$ 
pour des dur\'ees de survie possiblement censur\'ees $X_1^0,\ldots,X_n^0$.
On observe seulement $X_i=\min\{X_i^0,c_i\}$ et 
$\delta_i=I\{X_i^0\le c_i\}$ pour certains instants 
de censure $c_i$ qui sont donn\'es ou qui suivent 
une certaine loi de censure.
On \'etudie les questions suivantes:
Qu'est-ce que l'estimation de vraisemblance maxime et d'autres estimateurs 
estiment vraiment, quand la v\'eritable function de risque $\alpha(s)$
est diff\'erente des fonctions de risque param\'etriques?
Quelle est la loi limite d'un estimateur 
dans ces circonstances tellement dehors-le-mod\`ele?
Comment on peut assurer que des analyses traditionelles 
fond\'ees sur un mod\`ele sont `mod\`ele-robuste'?
Est-ce que le point de vue du mod\`ele agnostique 
sugg\'ere des abordages alternatives au probl\`eme de l`estimation? 
Quelles sont les cons\'equences de l'utilisation du la m\'ethode 
de Munchhausen fond\'e sur le mod\`ele et de celui qui est mod\`ele-robuste?
Comment est-ce qu'on peut faire la gen\'eralisation
des fonctions d'influence th\'eoriques et empiriques dans le cas
des donn\'ees censur\'ees?
Comment g\'eneraliser les m\'ethodes et r\'esultats 
quand on consid\`ere des mod\`eles plus complexes pour l'analyse des
donn\'ees de survie tels que 
mod\`eles de regression et des cha\^\i nes de Markof? 


\bye